\documentclass[
 aip,amssymb,jcp,
preprint,floatfix,
]{revtex4-2}
\usepackage{graphicx}
\usepackage{color}
\usepackage{hyperref}
\usepackage{float, enumitem}
\usepackage{array, multirow}
\usepackage{amsmath}
\usepackage{braket}
\usepackage{relsize}
\usepackage{newtxtext}
\usepackage{booktabs}
\usepackage{xcolor}
\usepackage{soul}
\usepackage{bbold}
\usepackage{url, dsfont}
\usepackage[T1]{fontenc}
\hypersetup{colorlinks,
    linkcolor={blue!75!black!80!yellow},
    citecolor={blue!75!black!80!yellow},
    urlcolor={blue!75!black!80!yellow}
    }

\begin{document}
\title{Designing robust molecular spins for quantum technologies with theoretical chemistry}

\author{Timothy J. Krogmeier}
\altaffiliation{These authors contributed equally to this work.}
\affiliation{Department of Chemistry, University of Minnesota, Minneapolis, MN 55455 USA}

\author{Pranay Venkatesh}
\altaffiliation{These authors contributed equally to this work.}
\affiliation{Department of Chemistry, University of Colorado Boulder, Boulder, CO, 80309, USA}

\author{Mikayla Z. Fahrenbruch}
\affiliation{Department of Chemistry, University of Minnesota, Minneapolis, MN 55455 USA}

\author{Anthony W. Schlimgen}
\affiliation{Department of Chemistry, University of Minnesota, Minneapolis, MN 55455 USA}

\author{Andr\'{e}s Montoya-Castillo}
\email{andres.montoyacastillo@colorado.edu}
\affiliation{Department of Chemistry, University of Colorado Boulder, Boulder, CO, 80309, USA}

\author{Kade Head-Marsden}
\email{khm@umn.edu}
\affiliation{Department of Chemistry, University of Minnesota, Minneapolis, MN 55455 USA}


\begin{abstract}
Molecular spins represent a versatile platform for quantum information science, with the potential to offer chemically tunable, addressable qubits. However, achieving this requires understanding and mitigating quantum decoherence. This Chapter provides a theoretical overview of current state-of-the-art chemical theory connecting \emph{ab initio} electronic structure with open quantum system dynamics to guide the rational design of long-lived molecular qubits. Beginning at the electronic level, multi-reference and relativistic electronic structure methods to parameterize effective spin Hamiltonians are discussed, with a primary focus on accurately capturing $g$-tensors, zero-field splitting, and hyperfine interactions. These parameters feed into models of spin-phonon and spin-spin coupling to quantify $T_1$ and $T_2$ relaxation across various environmental regimes. This Chapter evaluates a hierarchy of dynamical methods, ranging from factorization to matrix product state approaches, balancing computational cost against accuracy and generalizability. Ultimately, mapping these theoretical models to molecular architecture can establish design principles, such as isotopic substitution and spatial spin delocalization, to understand and extend coherence lifetimes.


\end{abstract}

\maketitle

\tableofcontents

\section{Background and Introduction}

\begin{figure}[b!]
    \centering
    \includegraphics[width=\linewidth, trim = 0cm 17.5cm 10.5cm 0cm, clip]{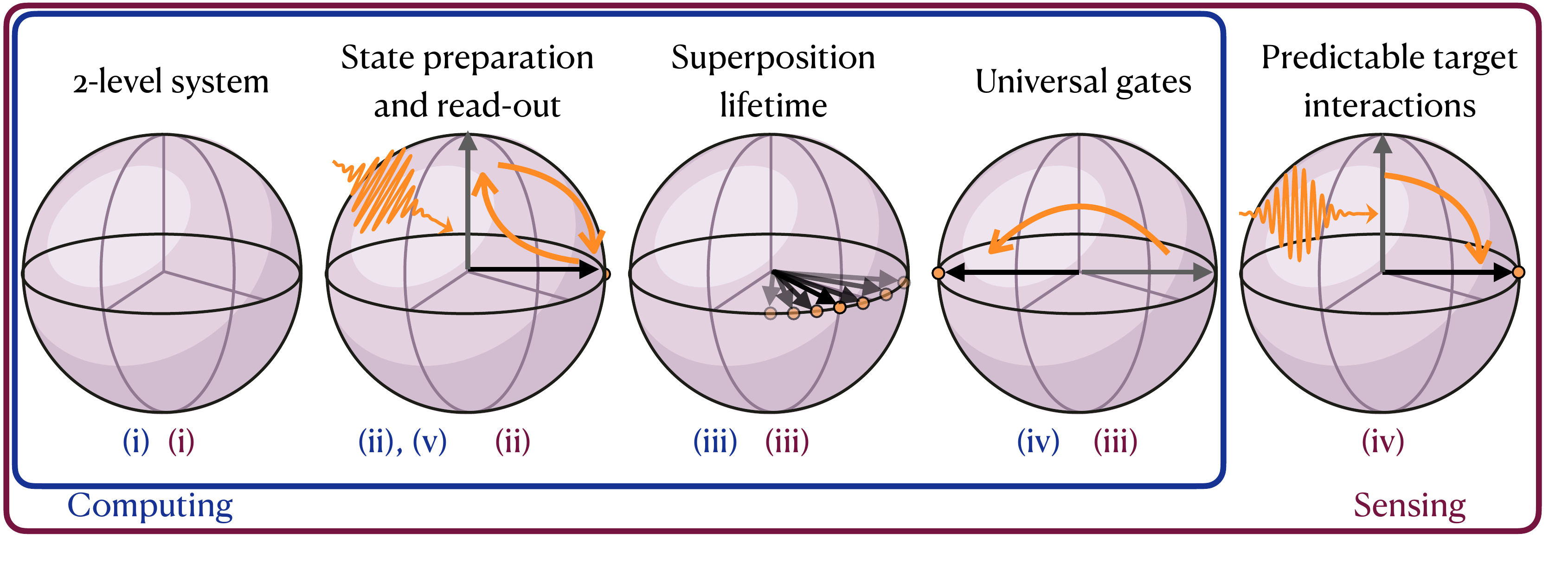}
    \caption{A schematic representation of the DiVincenzo (blue) and Degen (red) criteria for a quantum system to be a successful qubit or sensor, respectively.}
    \label{fig:degen-divincenzo-criteria}
\end{figure}

Emerging technologies based on quantum information science (QIS) promise to offer a route to efficient processing and storing of information beyond classical capabilities. Quantum computers, memories, and sensors are being developed from a variety of platforms, all aimed at exploiting quantum properties such as superposition and entanglement to improve the practical functionality of these devices. Early candidates in these quantum technologies include annealers,~\cite{Boixo:2014, Atanu:2022} superconducting circuits~\cite{Gambetta:2017, Wendin:2017, Citro:2024}, trapped ions for computing,~\cite{Clarke:2008,Gambetta:2017,Bruzewicz:2019} and defect centers for sensing.~\cite{Doherty:2013, Awschalom:2018, Wolfowicz:2021} While these are the most mature technologies to date, several other platforms are still developing, including neutral atoms and molecular spin systems.~\cite{Evered:2023, Gaita:2019,Wasielewski:2020, Albert:2020, Yu:2021} All of these technologies depend on the fundamental unit of quantum information: a quantum bit, or a qubit, which is effectively a two-level system that can be prepared in an arbitrary superposition state. Depending on the specific technological application, a quantum system must satisfy different criteria to be a good candidate. As one example, for processing information in quantum computation, the community frequently references DiVincenzo's criteria, which state that a quantum system must offer the following five traits: (i) discrete, resolvable energy levels; (ii) allow for state initialization; (iii) exhibit long coherence times; (iv) be addressable by a universal gate set; and (v) provide a state-specific readout mechanism.~\cite{Divincenzo:2000} For quantum sensing, the Degen conditions represent the analog: (i) discrete, resolvable energy levels; (ii) support the preparation and read-out of states; (iii) provide addressability of a coherence superposition state; and (iv) exhibit predictable interaction with a variable of interest.~\cite{Degen:2017} These properties are outlined schematically on Bloch spheres in Figure~\ref{fig:degen-divincenzo-criteria}.

\begin{figure}[h!]
    \centering
    \includegraphics[width=0.6\columnwidth, trim = 0cm 18cm 20cm 0cm, clip]{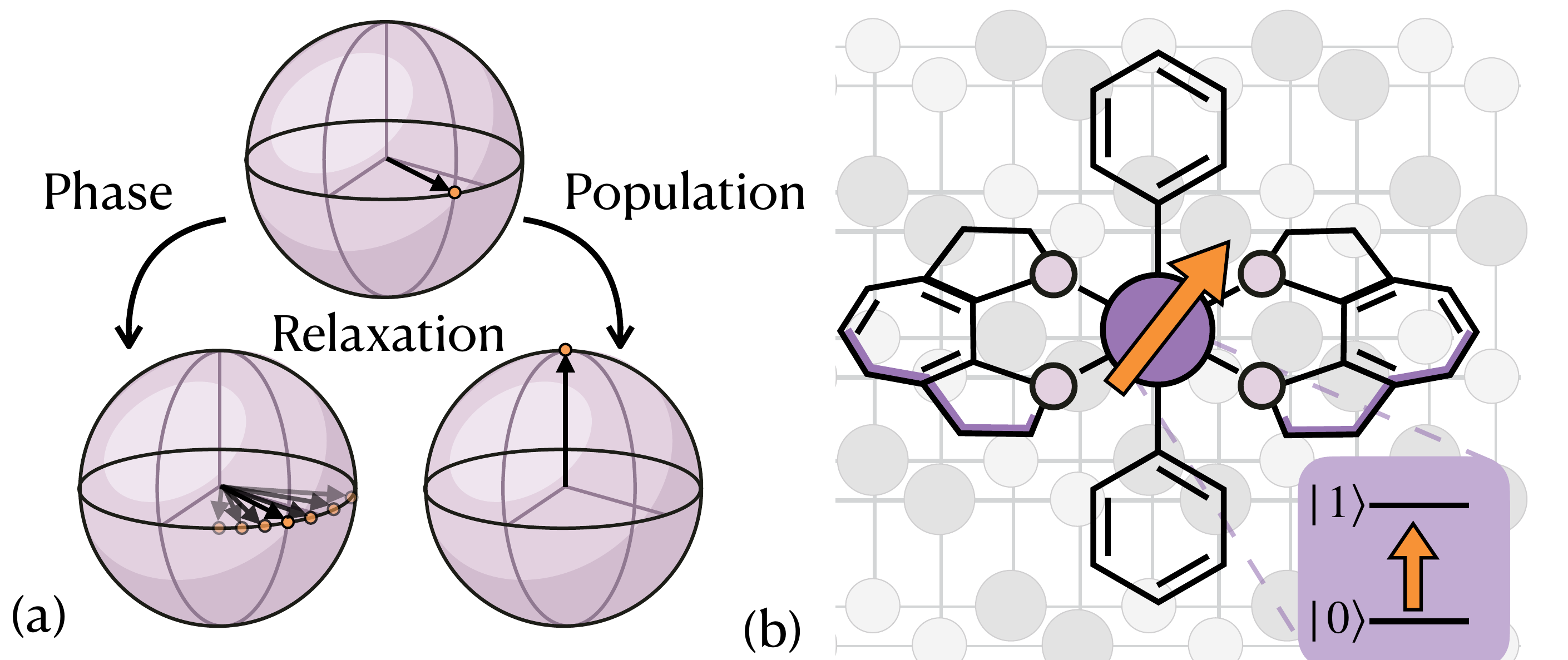}
    \caption{(a) Phase and population relaxation represented on a Bloch sphere of a qubit. (b) A molecular qubit candidate where a specific electronic spin state can offer the two-level system required by both DiVincenzo and Degen criteria. }
    \label{fig:molecular-qubit}
\end{figure}

A central challenge and opportunity facing QIS is that, by necessity, qubits are open systems that interact with their environment. These interactions are critical for technological functionality, but they also generally lead to phase and population relaxation. A qubit must be susceptible to external fields to enable state preparation, gate implementation, and readout, but this susceptibility exposes the qubit to environmental noise that limits its superposition lifetime. Generally, relaxation processes are categorized as either transverse, $T_2$, or longitudinal, $T_1$, relaxation, shown in Figure~\ref{fig:molecular-qubit}~(a). Transverse relaxation, commonly referred to as decoherence, defines the loss of phase coherence within the system density matrix. Taking the qubit basis to correspond to the eigenstates, $\{\ket{0}, \ket{1}\}$, of the $\sigma_z$ operator, decoherence corresponds to the decay of off-diagonal elements when the qubit state is parametrized on the Bloch sphere as,
\begin{equation}
    \ket{\psi} = \cos(\theta)\ket{0} + e^{i\phi}\sin(\theta)\ket{1}.
\end{equation}
Various mechanisms contribute to decoherence, including population relaxation along the $\theta$ direction, whose decay timescale is proportional to 2$T_1$ in the weak coupling and Markovian limits,~\cite{Ernst:1990} with deviations beyond these limits.~\cite{Chang:1993, Reichman:1996} Additionally, pure dephasing, associated with timescale $T_{\phi}$, and static inhomogeneities often referred to by timescale $T_2'$, both contribute to coherence loss.~\cite{Ernst:1990, Chang:1993, Abragam:2012} To experimentally obtain coherence or $T_2$ relaxation, dynamical decoupling in the form of applied pulse sequences is often used to minimize contributions from static inhomogeneities. For situations where dynamical decoupling is not used, often referred to as Free-Induction Decay (FID) or Ramsey experiments, the decoherence lifetime is referred to as $T_2^*$.\cite{Degen:2017}

Whether for quantum computing, communication, or sensing, chemistry plays a central role in understanding and controlling decoherence of molecular qubits and low-dimensional quantum materials. Built on decades of chemical intuition, mechanistic understanding of processes like charge transfer, singlet fission, intersystem crossing, excited-state spin evolution, and decoherence,~\cite{Zadrozny:2015, Shiddiq:2016, Atzori:2016, Amdur:2022} and advances in synthetic, metrological, and control capabilities, chemical tunability and control have enabled powerful demonstrations of two-qubit gates for controlled entanglement,~\cite{Ferrando:2016,  Mao:2023, Picard:2025} quantum memories built out of molecular magnets,~\cite{Leuenberger:2001, Mannini:2009} quantum teleportation for communication,~\cite{Rugg:2019} and molecular quantum sensor-enabled temperature metrology inside living cells.~\cite{Ishiwata:2026} These achievements required molecular qubit lifetimes in the microsecond to millisecond regimes,~\cite{Ardavan:2007, Timco:2009, Zadrozny:2015, Atzori:2016, Shiddiq:2016, Bayliss:2020, Maylander:2021, Dill:2023, Qiu:2023} and optical addressability.~\cite{Bayliss:2020, Chowdhury:2025, Feder:2025, Roggors:2025} 

Despite these advantages, several challenges currently limit the widespread use of molecular spins for quantum technologies. The most urgent of these include the ability to reliably measure the states at the individual qubit limit and identify and control the decoherence processes that hinder molecular implementations in realistic environments. To overcome these challenges, advances in two areas are critical: measurement and dynamics predictions. On the measurement side, there is a critical need for more precise and low-cost measurement protocols that are efficient and robust to experimental imperfections. On the theory and simulation side, the community needs efficient protocols to predict and analyze the quantum dynamics of \emph{ab initio}-parameterized many-spin Hamiltonians subject to external control protocols and interqubit interactions. Together, these two advances will offer the necessary capabilities to bring the promise of molecular QIS to fruition. Here, we focus on state-of-the-art methods for \emph{ab initio} prediction of molecular qubit dynamics, describe opportunities for theoretical and computational chemists, delineate basic phenomenology of molecular spin systems, include best practices for capturing electronic structure and relaxation dynamics of relevant materials, and discuss open challenges. In Section~\ref{ssec:e-struct} we provide an overview of current electronic structure theory approaches to computing molecular properties relevant for application in quantum computing and sensing. With these parameters, we transition to a discussion of theoretical methods to treat phase and population relaxation dynamics in Section~\ref{sec:dynamics}. We conclude in Section~\ref{sec:disc} with a discussion of remaining theoretical challenges and an outlook on the field. 

\section{Molecular qubits as dissipative open quantum systems}
\label{sec:context}

One description of the structure of a molecular qubit is through a general central spin Hamiltonian connected to a bath of spins and molecular vibrations and, when appropriate, lattice phonons,
\begin{align}
 \label{eq:ham}
    \hat{H}_{\rm T} = \hat{H}_{\rm q} + \hat{H}_{\rm s} + \hat{H}_{\rm q-s} + \hat{H}_{\rm ph},
\end{align}
with qubit (q), and spin bath (s), and vibrational and phonon (ph) contributions. The qubit Hamiltonian can be written as,
\begin{equation}
    \hat{H}_{\rm q} = \omega_q(\{ \mathbf{R}_{\alpha} \}) \hat{S}^z + \sum_{k,l\in\{x,y,z\}}\hat{S}^k \mathbf{D}^{k,l}(\{ \mathbf{R}_{\alpha}\}) \hat{S}^l \label{eq:Hq} ,
\end{equation}
where $\hat{S}$ represents the electron spin operators ($\hat{S}^x$, $\hat{S}^y$, and $\hat{S}^z$), ${\bf D}$ is the zero-field splitting tensor, $\omega_q = \mu_{B} g B_z$ is the spin's Larmor frequency, with $g$ the $g$-tensor of the electron, $B_z$ the magnitude of the magnetic field applied in the $z$ direction and $\mu_\mathrm{B}$ the Bohr magneton. The spin bath Hamiltonian is, 
\begin{equation}
    \hat{H}_{\rm s} = \sum_{i=1}^K\omega_i(\mathbf{R})\hat{I}_i^z + \sum_i \sum_{k,l\in\{x,y,z\}}\hat{I}_i^k \mathbf{Q}_i^{k,l}(\{\mathbf{R}_{\alpha}\}) \hat{I}_i^l + \sum_{i<j}\sum_{k,l\in\{x,y,z\}} \hat{I}_i^k \mathbf{J}_{i,j}^{k,l}(\{ \mathbf{R}_{\alpha}\})\hat{I}_j^l\label{eq:Hs},  
\end{equation}
where $K$ is the total number of nuclear spins considered, $\hat{I}$ represents nuclear spin operators ($\hat{I}^x$, $\hat{I}^y$ and $\hat{I}^z$), $\omega_i = \mu_n g_N B_z$ is the $i$th bath spin's Larmor frequency, ${\bf Q}$ corresponds to the quadrupole tensor of the nuclear spin environment, and $\mathbf{J}_{i,j}$ denotes the intra-bath coupling between the $i$th and $j$th spins. The phonon Hamiltonian is,
\begin{equation}
    \hat{H}_{\rm ph} = \sum_{\alpha=1}^N \frac{\mathbf{P}^2_{\alpha}}{2m_{\alpha}} + V(\{ \mathbf{R}_{\alpha}\}) \label{eq:Hph},
\end{equation}
where ${\bf P}$ represents momentum operators, $m_\alpha$ the mass of each atom, and ${\bf V}$ the potential. The qubit-spin bath interaction can be captured through, 
\begin{align}
\label{eq:sbham}
     \hat{H}_{\rm q-s} = \sum_{i=1}^K \sum_{k,l\in\{x,y,z\}} \hat{S}^{k} \mathbf{A}_{i}^{k, l}(\{\mathbf{R}_{\alpha} \}) \hat{I}_{i}^{l},
\end{align}
where ${\bf A_i}$ are the hyperfine tensors coupling the qubit and the $i$th bath spin. In the pure dephasing limit, which is generally applicable under sufficiently high magnetic fields and low temperatures that suppress spin flipping of the molecular qubit, only $k=z$ is included. Generally, one need not assume a harmonic or Gaussian description of the phonon Hamiltonian, $\hat{H}_{\rm ph}$; it may contain contributions from the motions of both atoms covalently bonded to the molecular qubit and those that compose the matrix in which it lies. While we did not include a separate qubit- or bath spin-phonon contribution to the Hamiltonian, the atomic positional dependence, $\{\mathbf{R}_{\alpha} \}$, of the spin parameters, $\{\omega_s, \mathbf{D}^{k,l}, \omega_i, \mathbf{Q}_i^{k,l},\mathbf{A}_{i}^{k, l}, \mathbf{J}_{i,j} \}$, indicates that one may expand these around a reference point to obtain terms that have no dependence on phonons, yielding a spin-only representation of the many-spin Hamiltonian, as well as those that contain first- and second-order contributions. 

At low temperatures, molecular qubit relaxation is primarily driven by magnetic dipole-dipole and hyperfine couplings between the qubit and surrounding environmental spins:~\cite{Zecevic:1998, Schweiger:2001} nuclear spins in the molecule and surrounding medium, and neighboring electron spins. In this limit, it is usually acceptable to adopt a spin-only Hamiltonian that neglects spin-phonon interactions or includes them only at the level of static disorder.~\cite{Canarie:2020} While spin-spin interactions dictate relaxation and dephasing in dense or concentrated spin environments, their contribution can be systematically mitigated through isotopic purification and magnetic dilution.~\cite{Bader:2014, Zadrozny:2015, Graham:2017} In addition, as temperature increases, atomic motions (vibrations and phonons) become thermally activated, modulating inter-spin distances and tensor orientations.~\cite{Lunghi:2019, Mirzoyan:2020} Thus, the influence of spin-phonon interactions depends intrinsically on temperature, since characteristic phonon energies are often comparable to or smaller than experimental thermal energies. For example, in the temperature range between 5 and 40~K, methyl groups on molecules or the host matrix can undergo quantum tunnelling, which introduces dephasing and spin relaxation to the central spin.~\cite{Jeschke:2022,Jahn:2024} Above 40~K, classical rotation of the methyl groups activate, limiting $T_2$ lifetimes in methylated molecular spins.~\cite{Kathirvelu:2009} As temperature approaches ambient conditions, phonon modes become thermally active and spin-spin relaxation is no longer the operative mechanism of total spin relaxation.~\cite{Atzori:2016, Lunghi:2020b} Furthermore, relaxation rates depend on the phonon density of states and its spectral overlap with spin-transition energies.~\cite{Garlatti:2020} 

Depending on the energy scales and parameter regime, different spin-phonon coupling channels can dominate longitudinal and transverse relaxation. These mechanisms are broadly categorized into direct, one-phonon processes, or indirect, two-phonon processes:
\begin{itemize}[noitemsep, topsep = 0pt]
    \item Direct Process: A single-phonon resonant transition directly matching the spin qubit operating frequency. These processes dominate $T_1$ relaxation at ultra-low temperatures and exhibit strong magnetic-field dependence.
    \item Raman Process: A non-resonant, two-phonon scattering process driven by virtual coupling to intermediate states.~\cite{Briganti:2021, Lunghi:2022} Raman processes typically dominate $T_1$ over low-to-intermediate temperature ranges.~\cite{Briganti:2021, Lunghi:2022}
    \item Orbach Process: A resonant mechanism where transitions between spin states proceed via a real, excited electronic state.~\cite{Orbach:1961} Depending on the theoretical framework, it can be described either as a sequential series of two single-phonon absorption and emission steps,~\cite{Orbach:1961, Briganti:2021} or as a resonant two-phonon scattering process.~\cite{Lunghi:2022} This process typically dominates both $T_1$ relaxation and $T_2$ dephasing at elevated temperatures.~\cite{Mondal:2022, Briganti:2025, Staab:2022}
\end{itemize}
Recently, the effect of three-phonon processes has also been evaluated:~\cite{Chanda:2026} these contribute only weakly to $T_1$ relaxation even at high temperatures, confirming the validity of the weak-coupling approximation when deriving generators for spin-phonon relaxation.

A central part of developing a dynamic understanding of decoherence in molecular qubits is constructing the appropriate Hamiltonian description. Evaluating the various contributions to the Hamiltonian in Eqs.~\ref{eq:ham} -- \ref{eq:sbham} requires state-of-the-art methods to determine spin parameters, spin-spin, and spin-phonon coupling tensors from electronic structure methods. When appropriate, one may alternatively resort to statistical models parameterized directly from experiment. A particularly useful and broadly adopted statistical model is the Gaussian limit. In the Gaussian limit, the dissipative dynamics of the qubit can be modeled by coupling to a continuum of harmonic oscillators that collectively represent the surrounding environment, including nuclear spins and phonons. This mapping reduces the full system-bath Hamiltonian to the spin-boson model---a system that generically describes decoherence in open quantum systems.~\cite{Caldeira:1981, Leggett:1987, Weiss:2012} Within this description, the microscopic details of the environment are encoded in a single function: the spectral density. Formally, the spectral density is defined as the Fourier transform of the bath's time-correlation function, which in the Gaussian limit specifies the entire statistics of the bath's fluctuations. In physical terms, the spectral density describes how strongly the qubit couples to environmental modes as a function of frequency, effectively acting as a fingerprint of the environment that drives decoherence.~\cite{Weiss:2012, Mukamel:1995, Boixo:2007, Schlosshauer:2019}

\section{Constructing the many-spin Hamiltonian}
\label{sec:struc}

Parameterizing the spin Hamiltonian in Eq.~\ref{eq:sbham} requires computing the zero-field splitting, $g$-, hyperfine, quadrupole, and spin-spin coupling tensors, along with their dependence on nuclear coordinates, for a given molecular qubit and its surrounding spin bath. These quantities can, in principle, be obtained from \emph{ab initio} electronic structure theory; however, the appropriate level of theory is sensitive to the character of the ground and low-lying excited states involved, ranging from well-behaved single-reference systems to strongly correlated, near-degenerate manifolds typical of transition-metal-based qubits. When such calculations are insufficiently accurate or unavailable, spectroscopic measurements offer a complementary route to parameterization. 
Such parameterizations recover an effective, statistical description instead of the full many-body Hamiltonian. 
Below, we outline the state-of-the-art theoretical and experimental methods used to extract these parameters in molecular spins relevant to QIS, beginning with electronic structure approaches to characterizing the spin-spin and spin-phonon interactions before turning to experiment-derived techniques.

\subsection{\emph{Ab initio} parameterization}
\label{ssec:e-struct}

Exploiting the potential of molecular qubits requires a detailed understanding of several aspects of molecular electronic structure. Though dependent on the experimental platform, it is typical to use the magnetic response of the electronic degrees of freedom to encode, transform, and read out quantum information. Common motifs for these species include transition metals or heavy elements, which are used to access the rich electronic structure in the $d-$ and $f-$electronic manifolds. While offering useful degrees of freedom, the choice of how to simulate these complex electronic states depends sensitively on the specifics of both the metal and its surrounding ligand environment. In this respect, the electronic densities, degenerate states, magnetic response, and relativistic effects increase the complexity of accurate simulations of many molecular qubit candidates. The appropriate electronic structure technique to simulate a given molecular qubit candidate is partially driven by the confluence of these effects.

For electronic ground states that are well-represented by a single-determinant wavefunction, density functional theory (DFT) provides a highly efficient and generally accurate description of the energy and density.~\cite{Hohenberg:1964aa,Kohn:1965aa} DFT is less reliable in the context of excited states partially because those states tend to be multireference, i.e. involving many determinants. Nonetheless, DFT is a highly valuable tool for describing ground-state energies, densities, and vibrational frequencies. In the context of molecular spin systems, DFT has been used recently to describe several molecular species,~\cite{Lunghi:2019, Garlatti:2020, Lunghi:2020c, Kazmierczak:2021, Vega:2022,Jahn:2024, Shushkov:2024, Mariano:2025} including a series of vanadium-oxo molecular qubits, where DFT is a reasonable model choice since the ground electronic states of the molecules are largely single reference.~\cite{Fataftah:2019, Chen:2020,Krogmeier:2024, Krogmeier:2026} The accuracy of properties that are computed at the DFT level for a single-reference electronic state is then dependent on the choice of basis set and functional. 

While highly scalable, standard DFT is ineffective for describing multideterminantal, or strongly correlated, wavefunctions. Instead, it is common to use multireference calculations with a complete active space (CAS) of correlated electrons and orbitals for strongly correlated electronic states.~\cite{Szalay:2012,Lischka:2018} Strong correlation can occur when different molecular orbitals become near-degenerate, resulting in several determinants with comparable weight in the wavefunction. For molecules containing transition metals or lanthanides, nearly-degenerate ground and excited states are common, which significantly complicates the modeling of these systems.~\cite{Pierloot:2011aa} Furthermore, the metal $d-$ or $f-$orbitals readily mix with ligand orbitals, which can cause delocalization, and the mixing can result in low-lying excited states. In these cases, finding even a qualitatively accurate wavefunction can be challenging, although recent progress in CI techniques has been vital in expanding the space of feasible calculations.~\cite{Schollwock:2005wz, Holmes:2016, Holmes:2017, Mazziotti:2012} CAS wavefunctions have recently been used to evaluate the excited-state manifold of molecular qubit candidates, revealing that careful treatment of the active space was required to correctly predict the molecular spectrum and spin delocalization.~\cite{Schlimgen:2023, Fataftah:2018aa, Sarkar:2025,Goh:2022} CAS methods have also been used to compute spin-phonon coupling properties~\cite{Lunghi:2022} and to characterize the crystal field in heavy element complexes.~\cite{Ungur:2017, Singh:2020} Additionally, DFT and multi-reference methods were compared in a review focusing on single-molecule magnets, which pose similar challenges in electronic structure modeling.~\cite{Atanasov:2015}

Unfortunately, these qualitatively accurate wavefunctions often fail to yield high-quality relative energies on their own. The correlation energy beyond the active space, often referred to as \emph{dynamic} correlation, plays an important role in providing quantitative accuracy in energies and properties. Reasonable quantitative accuracy can be achieved using second-order multireference perturbation theory (MRPT2);~\cite{Pulay:2011,Angeli:2001} however, the high cost of these calculations limits their routine use to chemical systems with perhaps 20-40 chemically relevant orbitals, while neglecting their environments.~\cite{Sarkar:2022} More recent developments have resulted in a combined approach using CAS self-consistent field (CASSCF) with DFT via multiconfigurational pair DFT (MCPDFT).~\cite{Bao:2020} Linearized PDFT (L-PDFT), the multistate variant of MCPDFT, has recently been extended to spin–orbit-coupled states and analytic dipole moments of ground and excited states, yielding accurate zero-field and fine-structure splittings.~\cite{Zhang:2025, Clifford:2025, Jangid:2026} Further, recent work using the PBE0@HF functional has been able to produce relative energies near CASPT2 levels of accuracy at mean-field cost.~\cite{Peng:2025}

The complexity of molecular qubit candidates presents a significant challenge to the simulation of their relevant properties. While DFT offers substantial advantages in computational cost, it is generally unreliable when considering cases involving strong correlation and degeneracy. Unfortunately, some techniques to prepare and interrogate molecular qubit states rely on the \mbox{(near-)degeneracies} present in the $d-$ or $f-$electronic manifolds, frequently rendering DFT an inappropriate model. Nonetheless, the recent advances in MCPDFT may provide a path forward for efficient calculation of these challenging ground and excited states. Wavefunction methods based on CASSCF are designed to handle degeneracies and state interactions consistently; however, their computational cost and required expertise frequently prohibit their routine application to complex molecular species. Here, unlike DFT, the cost of these calculations depends on the complexity of the electronic states under consideration. Furthermore, challenges arise due to the two-step process of generating a qualitatively accurate wavefunction with CASSCF, followed by perturbative corrections to the energy through MRPT2. Without the perturbative correction, the resulting energies and other properties of the ground and excited states are generally inadequate for accurate simulations of dynamics. These considerations lead to competing priorities when computing electronic properties, and suggest that careful probing of the electronic structure through various methods is likely required when generating parameters for dynamical simulations.

\subsubsection{Spin components}
\begin{figure}[h!]
    \centering
    \includegraphics[width=0.5\linewidth, trim = 0cm 17cm 29cm 0cm, clip]{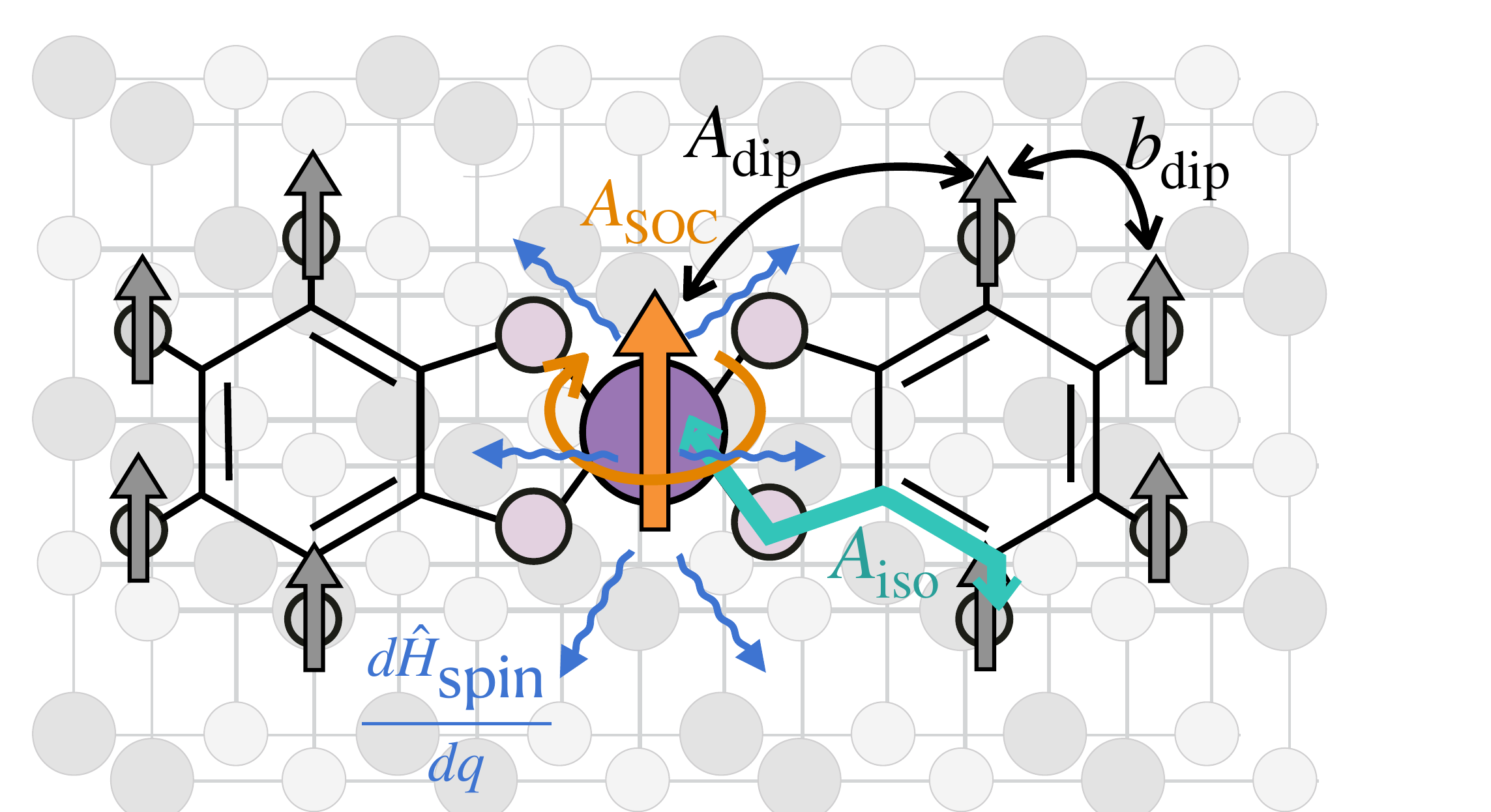}
    \caption{Electronic structure parameters required to understand and predict molecular properties necessary for use in QIS applications. These include hyperfine tensors (black arrows), dipolar couplings (black arrows), isotropic Fermi contact (teal arrow), spin-orbit coupling (orange arrow), and changes in the spin Hamiltonian due to interaction with phonon modes (blue arrows).}
    \label{fig:e-struc-paramts}
\end{figure}
The effective spin Hamiltonian, $\hat{H}_\mathrm{spin}$, contains only spin states and their interactions, connecting electronic structure to spin dynamics, and is obtained by integrating out spatial degrees of freedom from the full molecular Hamiltonian.~\cite{Kuprov:2023a} The spin Hamiltonian takes the following phenomenological form,
\begin{equation}
    \hat{H}_\mathrm{\rm spin} = \hat{H}_{\rm q}+\hat{H}_{\rm s} + \hat{H}_{\rm q-s},
    \label{eq:spinHamiltonian}
\end{equation}
where $\hat{H}_{\rm q}$ and $\hat{H}_{\rm s}$ are the Hamiltonians for the electron (qubit) and nuclear spins (spin bath) present in the system, respectively, and $\hat{H}_{\rm q-s}$ is the Hamiltonian describing interactions between the electron and nuclear spin degrees of freedom.~\cite{Kuprov:2023a} The electron spin or qubit Hamiltonian, $\hat{H}_q$, includes all terms coupling the electron spin to an external magnetic field $\overrightarrow{B}$ as well as the zero-field splitting in the case that $S > \frac{1}{2}$,
\begin{equation}
    \hat{H}_{\rm q} = \mu_\mathrm{B}\vec{B} g \hat{S} + \hat{S} {\bf D} \hat{S}.
\end{equation}
The $g$-tensor characterizes the interaction of the electron spin with an external magnetic field. The constant free electron $g$-factor is $g_e\approx 2.002319$; corrections to the full $g$-tensor can be computed as expectation values of the spin density and include relativistic mass corrections, spin-orbit coupling (SOC), and gauge correction terms.~\cite{Neese:2025} The relativistic mass correction is,
\begin{equation}
    g_\mathrm{RMC} = -\frac{\alpha^2g_e}{2S}\sum_{kl}\rho_{kl}\bra{\psi_k}\hat{T}\ket{\psi_l},
\end{equation}
where $S$ is the total spin, $\hat{T}$ is the kinetic energy operator, $\rho$ is the spin density, and $\alpha$ is the fine structure constant. SOC is often the dominant mechanism that shifts the $g$-tensor value, and is given by,
\begin{equation}
    g_\mathrm{SOC} = -\frac{g_e}{2S}\sum_{kl}\frac{\partial\rho_{kl}}{\partial \vec{B}}\bra{\psi_k}\hat{h}^\mathrm{SOC}\ket{\psi_l},
\end{equation}
where the partial derivative of the spin density $\rho$ is taken with respect to each Cartesian coordinate of the magnetic field vector, $\vec{B}$.~\cite{Neese:2025} Accurate computation of spin properties that are sensitive to spin-orbit coupling may require a careful choice of the model Hamiltonian, $\hat{H}_\mathrm{mol}$. Inclusion of SOC in electronic structure calculations is often based on perturbation theory, in which the zeroth order Hamiltonian is taken to be ``spin-free" and can include scalar-relativistic terms for higher accuracy, such as the Douglas-Kroll-Hess (DKH),~\cite{Hess:1985,Hess:1986,Douglas:1974,Jansen:1989,Wolf:2002,Reiher:2006,Nakajima:2012} zeroth-order regular approximation (ZORA) Hamiltonians,~\cite{Lenthe:1993,Lenthe:1994,Lenthe:1996} or exact two-component methods (X2C).~\cite{Franzke:2019,Pollak:2017} However, in cases where the SOC is strong, recent work has shown that inclusion of relativistic parameters in the zeroth-order Hamiltonian can improve the accuracy of computed spin dependent properties.~\cite{Lang:2019,Pathak:2017}

The zero-field splitting is prominent in molecular spin systems with $S > \frac{1}{2}$, in which the interaction lifts the degeneracy of the spin sublevels. In general, the zero-field splitting tensor, $D$, contains contributions from spin-spin interactions as well as spin-orbit coupling contributions. The first-order spin-spin contribution to the zero-field splitting is computed as an expectation value on the ground-state wavefunction, $\ket{\psi}$,
\begin{equation}
    D_{ss}^{\mu\nu} = \frac{1}{2}\frac{\alpha^2}{S(2S-1)}\bra{\psi}\sum_{i\neq j}\frac{|\vec{r}_{ij}|^2\delta_{\mu\nu} - 3(\vec{r}_{ij})_{\mu}\cdot(\vec{r}_{ij})_{\nu}}{|\vec{r}_{ij}|^5}\{ 2\hat{S}_{zi}\hat{S}_{zj}-\hat{S}_{xi}\hat{S}_{xj}-\hat{S}_{yi}\hat{S}_{yj} \}\ket{\psi},
    \label{eq:firstOrderZFS}
\end{equation}
where $i,j$ indexes electrons and $\mu,\nu$ indexes Cartesian coordinates $x,y$ and $z$. The second-order term includes a spin-orbit coupling correction.~\cite{Neese:2025} 

$\hat{H}_n$ is the spin Hamiltonian for nuclear spin $n$ and is given by,
\begin{equation}
    \hat{H}_n = \sum_n\mu_ng_\mathrm{N}\vec{B}\hat{I}_n + \hat{I}_n{\bf Q}\hat{I}_n + \sum_{n\neq m}\hat{I}_n{\bf J}_{nm}\hat{I}_m,
\end{equation}
where $\hat{I}_n$ is the operator for nuclear spin $n$, $\mu_n$ is the magnetic moment, $g_N$ is the isotropic g-factor for the nucleus, ${\bf Q} $ is the tensor describing the quadrupolar interaction for $I > \frac{1}{2}$, and $J_{nm}$ encodes the interaction between nuclear spins $n$ and $m$. The coupling between nuclear spins can be computed using the nuclear spin coordinates since they are well approximated by a point-dipole treatment,
\begin{equation}
    J_{nm}^{\mu\nu} = g_\mathrm{N,n}\mu_\mathrm{N,n}g_\mathrm{N,m}\mu_\mathrm{N,n}\frac{3\vec{r}_{nm\mu}\cdot\vec{r}_{nm\nu}-\delta_{\mu\nu}|r_{nm}|^2}{|\vec{r}_{nm}|^5},
    \label{eq:nuclearPointDipole}
\end{equation}
where $g_\mathrm{N,m}$ is the $g$-factor for nuclear spin $m$ with a gyromagnetic ratio indexed by $N$, and $\vec{r}_{nm}$ is the vector connecting nuclear spin $n$ to $m$. 

The quadrupolar interaction tensor, ${\bf Q} $, is an anisotropic property relevant for nuclear spins with $I>\frac{1}{2}$ that depends on the electric field gradient at a nucleus of interest, 
\begin{equation}
    {\bf Q} = \frac{1}{h}\frac{eQ_N}{2I_N(2I_N-1)}V,
    \label{eq:quadrupolarTensor}
\end{equation}
where $h$ is Planck's constant, $e$ is the fundamental charge, $I_n$ is the spin quantum number for nuclear spin $n$, $Q_n$ is the quadrupolar constant for nuclear spin $n$, and $V$ is the electric field gradient tensor which can be written as,~\cite{Stoll:2006} 
\begin{align}
    V(N) = -\sum_{kl}\rho_{kl}^e \bra{\psi_k} \frac{3\vec{r}_{ni,\mu}\cdot\vec{r}_{ni,\nu}-\delta_{\mu\nu}|r_{ni}|^2}{|\vec{r}_{ni}|^5} \ket{\psi_l} 
    + \sum_{m \neq n} Z_m \frac{3\vec{r}_{nm,\mu}\cdot\vec{r}_{nm,\nu}-\delta_{\mu\nu}|r_{nm}|^2}{|\vec{r}_{nm}|^5},
    \label{eq:efgTensor}
\end{align}
where $i$ indexes the electron, and $\rho_{kl}^e$ denotes the electron density. These parameters can be sensitive to the level and type of electronic structure theory.~\cite{Derevianko:2026} 

Finally, $\hat{H}_{\rm q-s}$ includes terms that drive electron spin relaxation and dephasing from nuclear spins,
\begin{equation}
    \hat{H}_{\rm q-s} = \sum_n\hat{S}{\bf A}_{n}\hat{I}_n,
\end{equation}
where ${\bf A}_{n}$ is the hyperfine tensor parametrizing the interaction between the electron and nuclear spins. The hyperfine coupling is often described with three terms,~\cite{Kuprov:2023a,Neese:2025} the first of which is the isotropic Fermi contact interaction,
\begin{equation}
    A_\mathrm{iso} = \frac{4\pi}{3}g_\mathrm{e}\mu_\mathrm{B}g_\mathrm{N}\mu_\mathrm{N}\rho(\vec{r}_{ni})(\langle \hat{S}_z \rangle)^{-1},
    \label{eq:Afc}
\end{equation}
where $g_e$ and $g_\mathrm{N}$ are the $g$-factor for the electron spin $e$ and nuclear spin $N$, respectively, $\mu_\mathrm{B}$ and $\mu_\mathrm{N}$ are the Bohr magneton and nuclear magneton, respectively, $\vec{r}_{ni}$ is a vector pointing from nuclear spin $n$ to electron spin $i$, and $\rho$ is the electronic spin density. $A_\mathrm{iso}$ is also referred to as the isotropic hyperfine coupling term, describing the magnetic interaction coming from the electron spin density located at nucleus $n$. For a localized electron spin, $A_\mathrm{iso}$ is likely a very small contribution for distal nuclear spins and can be neglected. However, in situations where hyperfine coupling to a transition metal nucleus with nonzero spin, $A_\mathrm{iso}$ may be very large and important to the dynamics. The second term contributing to the hyperfine coupling is the dipole-dipole coupling,
\begin{equation}
    A_\mathrm{dip}^{\mu\nu} = g_e\mu_\mathrm{B}g_\mathrm{N}\mu_\mathrm{N}\sum_{kl}\rho_{kl}\bra{\psi_k} \frac{3\vec{r}_{n\mu}\cdot\vec{r}_{n\nu}-\delta_{\mu\nu}|r_n|^2}{|\vec{r}_{n}|^5} \ket{\psi_l},
    \label{eq:Add}
\end{equation}
where $\rho$ is the spin density matrix, $\{ \ket{\psi} \}$ are the set of basis functions, and $\mu,\nu$ index Cartesian coordinates. This coupling describes the through-space interaction between the electron and nuclear spins, is highly anisotropic, and is important for accurate dynamics when some nuclear spins couple strongly to an unpaired electron. The third term is,
\begin{equation}
    A_\mathrm{SOC}^{\mu\nu} = -\frac{1}{2S}g_e\mu_\mathrm{B}g_\mathrm{N}\mu_\mathrm{N}\sum_{kl}\frac{\partial \rho_{kl}}{\partial \vec{I}_{\nu}}\bra{\psi_k}\hat{h}^\mathrm{SOC}_{\mu}\ket{\psi_l},
    \label{eq:Aso}
\end{equation}
which describes contributions to the hyperfine coupling from spin-orbit coupling of the electron and nuclear spin with operator $\hat{h}^\mathrm{SOC}$.~\cite{Neese:2025,Neese:2005} $\vec{I}_n$ is the vector of the nuclear spin magnetic moment operator, and the partial derivative is taken with respect to Cartesian coordinates $x$, $y$, and $z$. All three hyperfine terms require the spin density of the electron to be computed, and therefore require accurate wavefunctions. Furthermore, $A_\mathrm{SOC}$ depends on the manifold of electronic states and energies via the spin-orbit coupling operator $\hat{h}^\mathrm{SOC}$, requiring accurate excited-state energies, starting with a well-chosen basis set.  

Recent research has focused on benchmarking the use of DFT for calculating hyperfine tensors in different settings, including with relativistic corrections.~\cite{Cardona:2025} Additionally, DFT is often used in experiments to assign peaks in electron paramagnetic resonance experiments, calculate hyperfine couplings and $g$-tensor anisotropy, and predict electron spin densities.~\cite{Martins:2026,Bernier:2025} Multireference methods such as CASSCF have also been applied to obtain accurate energy splittings and ZFS parameters in a series of different molecular qubit candidates.~\cite{Jena:2026,Vega:2025,Baldinelli:2025} The anisotropy of the $g$-tensor has also been explored in molecular qubit candidates using CASSCF and $N$-electron valence second-order perturbation theory (NEVPT2).~\cite{Nath:2026} Additionally, the sensitivity of spin Hamiltonian parameters to correlation outside of an active space has been explored in the context of spin-phonon relaxation.~\cite{Haldar:2025} The level of electronic structure theory required will largely depend on the system of interest, although basis set selection always plays a critical role in accurate electronic structure. For light atoms such as hydrogen, which are common in molecular spin systems, the EPR-II basis sets give accurate hyperfine tensors at a relatively low cost.~\cite{Neese:2012,Barone:2008a,Barone:2008b,Barone:2008c,Barone:2010} First row transition metals are often well-described with the CP(PPP) basis set.~\cite{Gomez:2020,Neese:2002} 

\subsubsection{Spin-phonon couplings}

To parameterize spin-phonon Hamiltonians, one may approach the problem from a microscopic or a statistical perspective. From the microscopic perspective, one computes the derivatives of electronic spin parameters with respect to nuclear coordinates relative to a well-defined minimum-energy structure. By incorporating only low-order derivatives with respect to a reference structure, this approach is perturbative and appropriate only for fairly rigid molecules in a solid matrix where the harmonic approximation for describing nuclear motions is valid and the spin-phonon coupling is sufficiently weak to justify a low-order expansion of the couplings. In this picture, the spin-phonon coupling Hamiltonian $\hat{H}_\mathrm{s-ph}$ is derived by a Taylor expansion of system parameters with respect to normal mode coordinates $\alpha,\beta$ up to the second order,
\begin{equation}
    \label{eq:spin-phonon-hamiltonian}
    \hat{H}_\mathrm{s-ph} = \sum_\alpha \left(\frac{\partial \hat{H}_\mathrm{spin}}{\partial q_\alpha} \right)_0 q_\alpha + \sum_{\alpha \geq \beta}\left(\frac{\partial^2\hat{H}_\mathrm{spin}}{\partial q_\alpha \partial q_\beta}\right)_0 q_\alpha q_\beta ,
\end{equation}
where $q_{\alpha}$ denotes the $\alpha^{\rm th}$ phonon mode obtained from a second-order expansion around equilibrium geometries.~\cite{Escalera-Moreno:2017,Lunghi:2022} Calculating the spin-phonon coupling coefficients is a computationally demanding task, and here we briefly outline the state-of-the-art computational methods.

From the statistical perspective, one aims to capture the spin components of the thermally accessible average structure while accounting for Gaussian fluctuations of spin Hamiltonian parameters where nuclear motions of the molecule and the host matrix modulate the electronic gyromagnetic $g$, hyperfine ${\bf A}$,~\cite{Lunghi:2019} and zero-field splitting ${\bf D}$ tensors~\cite{Hay:2019, Lunghi:2022} by altering spin-orbit coupling and local geometry.~\cite{Lunghi:2019} This approach is valid where the central limit theorem applies and has, to date, only been used to compute one-phonon processes. These modulations induce pure dephasing, similar to spin-spin interactions; however, in the high-field limit, fluctuations in $A_{n}$ with respect to these nuclear modes are often neglected.~\cite{Albino:2019, Escalera-Moreno:2017, Dmitriev:2025, Aruachan:2023, Aruachan:2026}

The finite difference method serves as a primary numerical tool to determine spin-phonon coupling elements, where electronic structure calculations, such as DFT or CASSCF, are repeated over systematic molecular or unit-cell distortions. Early approaches involved distorting molecular geometries and fitting electronic structure parameters to polynomials to extract first- and second-order derivatives,~\cite{Escalera-Moreno:2017} with additional identification of specific relaxation pathways via group theory selection rules.~\cite{Kazmierczak:2021} These methods evolved into workflows combining periodic DFT and CASSCF to account for lattice anharmonicity.~\cite{Lunghi:2017a} While accurate, finite difference methods are computationally expensive, often requiring thousands of high-level electronic structure calculations. For example, 1830 CASSCF calculations were needed to determine the spin-phonon coupling operator for a single dysprosium molecule.~\cite{Briganti:2021} 

Analytical derivative techniques, such as Density Functional Perturbation Theory (DFPT) and analytical CASSCF gradients,~\cite{Baroni:2001, Escalera-Moreno:2017} circumvent the high computational cost of grid-based finite displacements and have emerged as powerful alternatives. These methods compute spin-phonon coupling tensors directly from the molecular or crystalline Hessian at equilibrium. Recent work employed analytic derivatives of crystal-field parameters via a diabatic representation of the Hamiltonian to build linear vibronic Hamiltonians, i.e., those containing only one-phonon coupling terms.~\cite{Staab:2022} These approaches evaluate state mixing due to nuclear displacements around an equilibrium geometry and can be combined with molecular dynamics trajectories to sample spin Hamiltonian parameters across thermally accessible conformations.~\cite{Alnami:2023, Thompson:2025} However, there are significant challenges in developing and implementing many-electron theories with analytical derivatives, which has limited their practical application. 

To increase computational efficiency when analytical gradients are unavailable, one can lower the cost of repeated electronic structure calculations by training machine-learned interatomic potentials whose cost of inference is significantly smaller than that of solving the electronic Schr\"odinger equation from scratch. Such approaches have already proven profitable,~\cite{Lunghi:2020a,Lunghi:2020c} and CASSCF- and DFT-based machine-learned interatomic potentials have been trained to determine the response of the axial anisotropy term to phonon modes for a high-spin cobalt complex.~\cite{Lunghi:2020d, Lunghi:2022} A more recent application of machine-learned interatomic potentials to cobalt and dysprosium compounds in the gas phase achieved good agreement with the traditional finite differences method while reducing computational cost by 80\%.~\cite{Briganti:2025} 

The statistical approach, where one samples the ground electronic state potential energy surface at finite temperature often with \emph{ab initio} accuracy to interrogate the statistical fluctuations of theoretically or experimentally interesting quantities, has a long history in physical chemistry, \cite{Chandler:1993, Makri:1999, Limmer:2024} including in treatments of solvation,\cite{Chandler:1993} charge transfer,\cite{Song:1996, Georgievskii:1999, Coello:2024, Mayers:2018, Limmer:2013, Willard:2013} electrochemistry,\cite{Uralcan:2016, Kuharski:1988, Coello:2024} density fluctuations in liquids,\cite{Andersen:2005, Hansen:2006} and spectroscopic responses of condensed phase systems.\cite{Mukamel:1995, Lee:2016, Loco:2018, Zuehlsdorff:2019, Wiethorn:2023} Applied in a QIS context, the statistical approach leverages molecular dynamics simulations to either sample static disorder or parameterize the dynamic disorder of the spin tensor for molecular qubits embedded in glass matrices and crystals. \cite{Martin:2019, Jahn:2024} In the static limit, it is common to use frames from molecular dynamics simulations as samples of various realizations of static disorder and average over the resulting coherence dynamics of these disorder realizations.~\cite{Canarie:2020, Jahn:2024} From the dynamic perspective, molecular dynamics simulations are used to compute the spectral density associated with $g$-tensor fluctuations.~\cite{Lunghi:2020c, Aruachan:2026} Just as the microscopic approach implicitly assumes the validity of harmonic phonons and the validity of low-order perturbation theory in capturing spin-phonon couplings, statistical approaches imply a different set of assumptions and approximations. For example, when focusing only on static disorder, the assumption of extreme timescale separation is necessary.\cite{Mukamel:1995} For dynamic disorder captured via spectral densities, a (semi)classical approximation is often adopted that replaces the quantum equilibrium time correlation function required to extract the spectral density with its classical counterpart, usually modified by a quantum correction factor that enforces the symmetries of the real and imaginary parts of the quantum correlation function.\cite{Zuehlsdorff:2019, Wiethorn:2023, Egorov:1999, Craig:2004, Ramirez:2004} The applicability of the statistical approach for dynamic disorder is rooted in the applicability of the central limit theorem, often ensured by the fact that a large number of atoms in the condensed phase weakly modulate the electronic and spin parameters,\cite{Chandler:1993, Makri:1999, Limmer:2024} leading to Gaussian statistics for their fluctuations around equilibrium. To date, dynamic, or frequency-resolved, spin-phonon interactions in QIS have been captured only at linear order and at the level of lattice-induced fluctuations of the parametrized gyromagnetic tensor calibrated with experimental $T_1$ measurements.~\cite{Aruachan:2023, Lunghi:2020c} Here, the influence of the environment on the system Hamiltonian is partitioned into fluctuations of the $g$-tensor and of the local magnetic field, allowing for simultaneous calculation of $T_1$ and $T_2$ times as a function of magnetic field strength and temperature. These statistical treatments, however, struggle with the parameterization of two-phonon interactions responsible for Raman- and Orbach-type relaxation.

\subsection{Experiment-derived parameterization}
\label{ssec:experiment-struct}

Spectroscopic measurements have traditionally been used to build and parameterize simple models of quantum relaxation. These have a strong tradition in chemistry and physics. Here, we focus our discussion on spectroscopic techniques and protocols that can be used to infer spin-spin and spin-phonon couplings in molecular systems and QIS applications. The simplified nature of the model that often underlies such parameterizations is not a full many-spin Hamiltonian with spin-phonon couplings; instead, these parameterizations generally recover simple spectral densities appropriate for the Gaussian limit. Thus, in this limit, all spin-spin and spin-phonon interactions become indiscriminately mapped onto bilinear coupling to a bosonic environment, recovering the form of a spin-boson model. 

In QIS, a primary workhorse for extracting these spectral densities is quantum noise spectroscopy.~\cite{Suter:2016, Szankowski:2017, Norris:2016, Vezvaee:2024, Shitara:2026} This involves measuring a qubit's coherence in the pure dephasing limit or performing full tomography in the full decoherence limit. While these methods primarily work in the Gaussian limit,~\cite{Krzywda:2019} recent advances have facilitated the generalization into the non-Gaussian regime.~\cite{Ramon:2015, Norris:2016, Ramon:2017, Sung:2019} Nevertheless, the measurements and error thresholds required for successful extraction of Hamiltonian couplings beyond the Gaussian limit are exacting and have only been achieved in macroscopic superconducting qubits,~\cite{Sung:2019, Vepsalainen:2022} but not yet in molecules. 

Estimating the spectral density from coherence measurements can be approached in several ways. Dynamical decoupling noise spectroscopy (DDNS) uses long, periodic pulse trains, whose filter functions approach a frequency comb that samples the spectral density at specific frequencies.~\cite{Alvarez:2011} By repeating the measurement at a set of commensurate pulse spacings, one obtains a set of decay rates that can be inverted for the spectral density at those discrete frequencies. Fourier transform noise spectroscopy (FTNS) instead inverts the filter-function integral analytically for the free induction decay and spin-echo cases, showing that the spectral density is the Fourier transform of the second time-derivative of the log-coherence.~\cite{Vezvaee:2024} This gives a one-to-one map between the spectral density and the coherence, but the double derivative demands finely sampled and well-averaged noise---generally cumbersome requirements when performing measurements in the individual qubit limit. Alternatively, one can view this as an inverse problem that can be solved by optimizing a variational basis set, as in the case of variational quantum noise spectroscopy (VQNS).~\cite{Shitara:2026} In VQNS, a trial spectral density built from an overcomplete basis is refined until the coherence curves it predicts reproduce those measured under several pulse sequences at once. Since the optimization starts from randomly sampled initial parameters, repeated runs also quantify the uncertainty in the reconstruction and indicate which further experiments would minimize reconstruction error. Only recently has there been a proof-of-principle extension of this noise spectroscopy problem to the full decoherence limit,~\cite{Wise:2021, Gupta:2025} albeit with extensive and expensive measurement requirements and temporal resolution. While noise spectroscopy elucidates the frequency ranges over which a dynamical decoupling sequence is expected to eliminate contributions to decoherence, this frequency-by-frequency characterization breaks down beyond the Gaussian limit where combination bands and overtones contaminate the spectral density.

In physical chemistry, the primary concern for many decades has been to infer exciton-vibrational couplings from experiment. Here, simple frameworks like the Brownian oscillator or Frenkel exciton models aim to encode solvation effects in the photochemistry of polymers,\cite{Spano:2006} pigment-protein complexes,\cite{Jang:2018, Curutchet:2017, Jha:2026} and other biologically relevant chromophores.~\cite{Improta:2016, Gozem:2017, Sala:2018} The spectral density can be experimentally inferred via techniques such as photon echos,\cite{Fleming:1996, Cho:1996, Cheng:2007, Pachon:2014, Jeon:2019} resonance Raman,~\cite{Gustin:2023, Kazmierczak:2024, Gustin:2025} fluorescence line narrowing,~\cite{Adolphs:2006} and two-dimensional electronic spectroscopies.~\cite{Parker:2022, De_wit:2025} There remains an open problem in how to leverage these techniques to parameterize the spectral density experienced by substates in the magnetic manifold of electronically excited states of molecular qubits, like single-molecule magnets and singlet fission materials. 

To characterize molecular qubits from experiment and build a statistical description of their decoherence requires bridging the capabilities of traditional QIS approaches and physical chemistry. In particular, it is unclear how physical chemistry techniques that aim to recover exciton-phonon coupling exclusively, and sampled at optical frequencies, can distinguish the contributions from a densely populated magnetic manifold that forms the focus of the molecular qubit implementation. In addition, the physical chemistry techniques reported operate at the level of ensemble measurements, whereas it is often important to characterize the decoherence of qubits on an individual basis. Because quantum noise spectroscopies address the individual-qubit limit, these offer a potentially fertile path to molecular qubit characterization. However, these also face challenges as they often require the ability to localize a single qubit in a solid matrix. As a compromise between these two approaches, recent work leverages time-resolved and dynamical decoupling-based EPR measurements that aim to elucidate the relaxation mechanisms of the magnetic manifold of electronically excited of molecules (metal complexes and singlet fission materials) where qubits and their multilevel generalization may be implemented.~\cite{Zadrozny:2015, Yu:2016, Fataftah:2019, Dill:2023, Reid:2024, Nelson:2017, Wu:2018, Smyser:2020}

\section{Predicting the quantum dynamics of decoherence}
\label{sec:dynamics}

\begin{figure*}[h!]
    \centering
    \includegraphics[width=0.75\linewidth]{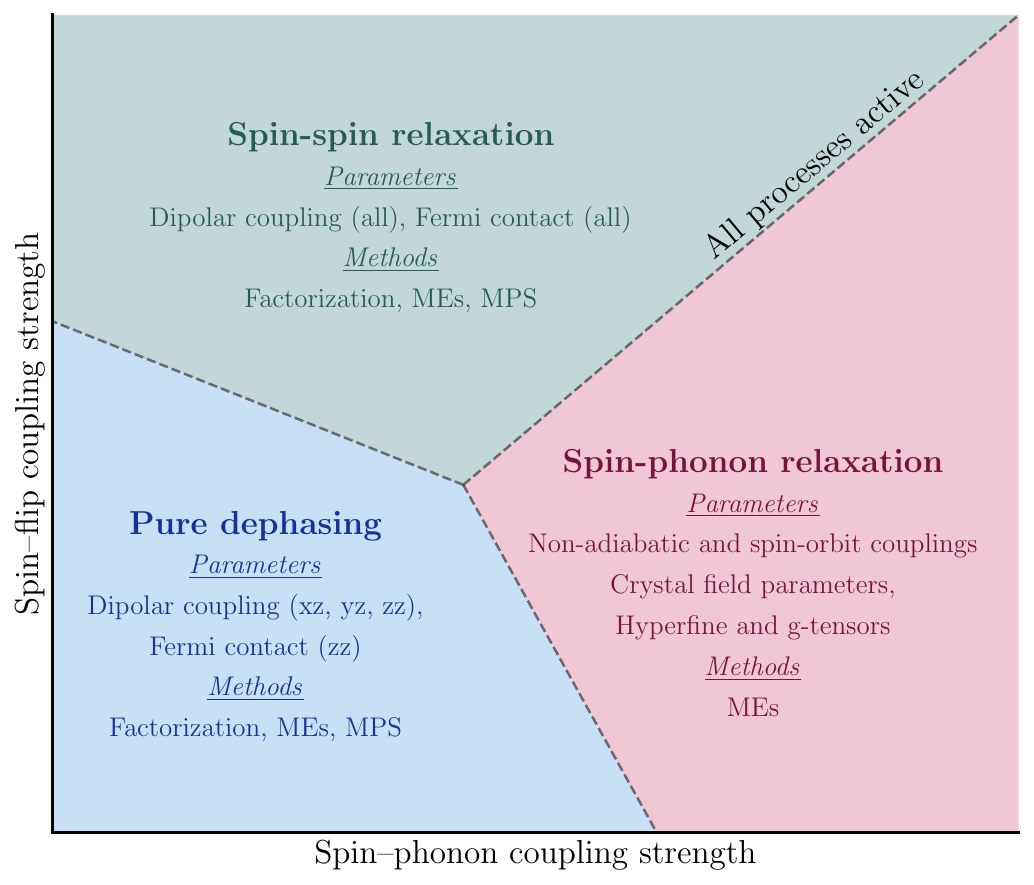}
    \caption{Schematic of spin relaxation regimes and primary theoretical methods as a function of spin-flip and spin-phonon coupling strengths. Green: Regime where spins are concentrated and non-secular dipolar and Fermi contact terms drive spin flips and dephasing. Blue: Regime where spins are dilute and dephasing from nuclear spin bath fluctuations dominates. Red: Thermally active regime dominated by spin-phonon relaxation via vibrational modulation of spin parameters. Theoretical methods include the broad categories of factorization approaches, Matrix Product State (MPS) methods, and Master Equations (MEs).}
    \label{fig:Spin-dyn-schem}
\end{figure*} 
Molecular spin lifetimes are governed by distinct relaxation channels that dominate across different physical regimes, as summarized schematically in Figure~\ref{fig:Spin-dyn-schem}. In concentrated ensembles or systems with strong inter-spin interactions, electron spin flips contribute appreciably to population and phase relaxation of electron spins, shown in the green region of Figure~\ref{fig:Spin-dyn-schem}.~\cite{Schweiger:2001, Ardavan:2007, Takahashi:2011} Diluting the molecular spin species and applying moderate-to-high magnetic fields suppresses spin flips of the central electron spin.~\cite{Takahashi:2011, Wedge:2012, Bader:2014} In the dilute limit, the electron relaxation dynamics are dominated by pure dephasing driven by nuclear spin bath fluctuations via secular and pseudo-secular dipolar and Fermi contact hyperfine interactions, shown in the blue region of Fig.~\ref{fig:Spin-dyn-schem}.~\cite{Yang:2008, Canarie:2020, Jahn:2024} The red area of Fig.~\ref{fig:Spin-dyn-schem} depicts the onset of spin-phonon relaxation, driven by both adiabatic modulation of spin parameters and non-adiabatic coupling where electronic and nuclear motion can no longer be fully separated.~\cite{Lunghi:2019, Kazmierczak:2021, Shushkov:2024} In molecular spins, vibrational modes and acoustic phonons modulate the zero-field splitting, $\mathbf{g}$-tensor, and hyperfine tensors. Furthermore, non-adiabatic spin-vibrational coupling and state mixing induce longitudinal and transverse relaxation, a regime that typically correlates with increasing temperature and lower-lying electronic excited states. Across all regimes, the many-body dynamics of the environmental spin bath are important to the relaxation or dephasing rate of the central spin. $T_2$ dynamics in the blue and green regions of Fig.~\ref{fig:Spin-dyn-schem} in molecular spins are sensitive to the number, configuration, and type of bath spins present in a molecule, meaning that accurate modeling of a specific molecular spin requires careful treatment of all relevant processes. 

To model these diverse mechanisms, there are various theoretical frameworks tailored to specific parameter regimes: Cluster Correlation Expansion (CCE),~\cite{Onizhuk:2021, Yang:2008, Yang:2009} Matrix Product State (MPS),~\cite{Li-SBt:2026, Lindoy:2018} and Master Equation (ME) methods excel in capturing pure dephasing and nuclear bath correlations,\cite{Krogmeier:2026, Onizhuk:2024a} while MEs currently address spin-phonon relaxation.\cite{Lunghi:2022, Lunghi:2020b,Reta:2021, Briganti:2021, Lunghi:2019, Lunghi:2026, Evans:2019, Goodwin:2017, Yu:2020} Additionally, since spin relaxation is measured in large ensembles of molecular spin systems, inhomogeneity of the ensemble and the applied magnetic field also contribute to the spin relaxation rate.\cite{Eaton:2000, Wasielewski:2020} Dynamical decoupling sequences can be used to selectively suppress both static and some dynamic contributions to the relaxation rate.~\cite{Uhrig:2007, Alvarez:2011, Suter:2016, Cywinski:2008} In the Gaussian limit, the filter function formalism elucidates the frequency ranges over which a dynamical decoupling sequence can eliminate the contribution of fluctuations to decoherence.~\cite{Krzywda:2019, Szankowski:2017, Shitara:2026} Accurate theoretical modeling must therefore match both the operational mechanism and the specific experimental pulse sequence. Thus, an ideal dynamics method should deliver the ability to accurately and efficiently predict the quantum dynamics of many-spin Hamiltonians including spin-spin interactions, the influence of static environmental heterogeneity and applied fields, the presence of one- and two-phonon coupling mechanisms, and the impact of arbitrary driving, such as dynamical decoupling sequences. Below, we outline state-of-the-art theoretical and computational methods for predicting spin-spin and spin-phonon relaxation dynamics in molecular quantum technologies.

\subsection{Exact Diagonalization}

Exact diagonalization offers a direct path to simulating the quantum dynamics of a many-spin Hamiltonian. It makes no approximation to the structure of the many-body Hilbert space or the dynamics and hence retains the full system-environment density matrix,~\cite{Sandvik:2010, Weinberg:2017} making it a natural benchmark against which other numerically exact and approximate methods can be validated. This fidelity comes at a steep price: the dimension of the many-body Hilbert space grows exponentially with the number of spins, which restricts exact diagonalization to comparatively small spin systems.

A composite system-environment state evolves in time through unitary propagation, 
\begin{equation}
    \rho_\textrm{T}(t) = \hat{U}^{\dagger}(t)\rho_\textrm{T}(0)\hat{U}(t),
    \label{eq:unitary-evol}
\end{equation}
or with the Liouville-von Neumann equation in atomic units,
\begin{equation}
    \frac{d\rho_\textrm{T}(t)}{dt} = -i [\hat{H}_{\textrm{spin}}(t), \rho_\textrm{T}(t)]
\end{equation}
where $[\cdot, \cdot]$ is the commutator. In both of these expressions, $\rho_\textrm{T}(t)$ is the total composite density matrix of the system and the environment. The reduced density matrix of the system of interest can be obtained through a partial trace procedure and is denoted by $\rho(t)$. In the RDM, the diagonal elements, $\rho_{ii}(t)$, provide information on populations, while the off-diagonal elements, $\rho_{ij}(t)$, represent coherences between states. In general, this Hamiltonian is time-dependent due to molecular motion and the application of dynamical decoupling.~\cite{Uhrig:2007} Assuming a low enough temperature such that decoherence and molecular motion occur on different timescales, approximately below 100 K,~\cite{Albert:1972} the time dependence of the Hamiltonian is entirely due to dynamical decoupling. The theoretical treatment of pulse sequences simulates experiments that measure spin echoes, such as the Hahn-echo experiment,~\cite{Hahn:1950} in which a single $\pi$-pulse is applied halfway through the evolution of the spin system. Additional $\pi$ pulses can be included to extend spin lifetimes. Pulse sequences with more than one, equally spaced $\pi$ pulse are referred to as Carr-Purcell-Meiboom-Gill pulse sequences~\cite{Carr:1954, Meiboom:1958}, or CPMG-$n$ pulse sequences, where $n$ denotes the number of pulses included in the sequence. Pulse sequences are employed to measure $T_2$ without complications from $T_2^*$ dephasing due to static noise sources, isolating spin relaxation due to dynamic fluctuations of the environment. For spin systems, at typical experimental energy scales of $k_BT \sim 10^{12}$~Hz, or for temperatures above 5 K, it is common to invoke the infinite-temperature approximation for the nuclear and electronic spins in the bath, since the energy scales of many-spin Hamiltonians generally span microwave frequencies. 

One can evaluate the exact time-evolution operator $\hat{U}(t)$ for a composite system-environment by diagonalizing the Hamiltonian. The evolution of the density matrix is then obtained through Eq.~\ref{eq:unitary-evol}; however, since the Hamiltonian scales exponentially with the size of the system, an exact diagonalization calculation is limited to around 20 spins.~\cite{Sandvik:2010} This cost can be ameliorated by using Krylov subspace methods that do not rely on diagonalizing the entire Hamiltonian,~\cite{Weinberg:2017, Weinberg:2019, Prelovvsek:2013} allowing for dynamics of larger spin systems. However, the calculation still scales exponentially with system size and hence computing spin dynamics with hundreds of spins is unfeasible without further approximations. As such, exact diagonalization is best suited to describe spin relaxation and dephasing due to nearby electron or nuclear spins in small molecular spin systems at low temperatures. This could be a suitable model for an isolated molecule with an electron and molecular nuclear spins, or several nuclear spin-free molecules dense enough that electron-electron spin interactions contribute significantly to relaxation.

While exact diagonalization can, in principle, be used to study decoherence in the presence of spin-phonon coupling,~\cite{Alvermann:2009} it requires converging the dynamics with respect to the number of states retained from the infinite Hilbert space of each phonon. As temperature increases, this number becomes ever larger. Unfortunately, the dimensionality grows exponentially with system size, number of phonons, and the number of states retained from their Hilbert space, making it practically unfeasible for systems with a realistic number of phonons. 

\subsection{Factorization Approaches}

The exponential scaling of exact diagonalization arises because of the growing dimensionality of the many-body Hilbert space with increasing numbers of spins; however, factorizing, or clustering, environment spins can sometimes accurately describe the dynamics in molecular systems. In recent years, the cluster-correlation expansion (CCE) method has become the state-of-the-art method for simulating coherence dynamics for spin systems with interacting baths.~\cite{Yang:2008, Yang:2009} In the CCE formulation, the coherence function is expressed as a product of irreducible terms from combinations, or sub-clusters, of bath spins,
\begin{align}
    \label{eq:cce-prod-exp}
    C(t) = \prod_A \tilde{C}_A(t) = \prod_i \tilde{C}_{\{i\}}(t) \prod_{i,j} \tilde{C}_{\{ij\}}(t)...,
\end{align}
where $\{i,j,\cdots\}$ denote all possible subclusters of the total spin bath that include spins indexed by $i, j, \cdots$. That is, $\{i\}$ denotes spin subclusters consisting of one spin, $\{ij\}$ denotes spin subclusters consisting of two spins, and so on. The contribution from individual clusters is given by,
\begin{align}
    \tilde{C}_A(t) = \frac{C_A(t)}{\prod\limits_{B\subset A} \tilde{C}_B(t)}.
\end{align}
The coherence function for clusters is computed by evaluating,
\begin{align}
    C_A(t) = \langle 0 | \hat{U}_A(t) \rho(0) \hat{U}_A^{\dagger}(t) | 1 \rangle,
    \label{eq:cce-cluster-contribution}
\end{align}
where the time-propagator for cluster $A$, corresponding to sub-system Hamiltonian $\hat{H}_A$, subject to a CPMG-$n$ pulse sequence separated by time delay $\tau$ and characterized by rotational angle $\phi$, is given by,
\begin{equation}
    \label{eq:timeprop}
    \hat{U}_A = \mathcal{T} \left[ {\rm e}^{-\frac{i}{\hbar} \hat{H}_A \tau} {\rm e}^{-\frac{i}{\hbar} \hat{S}_y \frac{\phi}{2}} {\rm e}^{-\frac{i}{\hbar} \hat{H}_A \tau}\right]^n,
\end{equation}
where one obtains the cluster Hamiltonian $\hat{H}_A$ by first projecting the Hamiltonian on each central spin state $\alpha = \ket{0}, \ket{1}$,
\begin{equation}
    \hat{H} = \lvert 0 \rangle \langle 0 \lvert \otimes \hat{H}^{(0)} + \lvert 1 \rangle \langle 1 \lvert \otimes \hat{H}^{(1)},
\end{equation}
then tracing out the bath spins excluded from the cluster,
\begin{equation}
    \hat{H}_A^{(\alpha)} = \sum_{i \in A} \langle \alpha \lvert \hat{S} \lvert \alpha \rangle {\bf A}_i \hat{I}_i + \sum_{i \in A} (\hat{I}_i {\bf Q}_i \hat{I}_i + B \gamma_i \hat{I}_i) + \sum_{i,j \in A} \hat{I}_i {\bf J}_{ij} \hat{I}_j + \sum_{i \in A, j \notin A} \hat{I}_i {\bf J}_{ij} \langle \hat{I}_j\rangle.
    \label{eq:clusterHamiltonian}
\end{equation}
$\langle \hat{I}_j\rangle$ indicates an average over all nuclear spins outside of the cluster $A$, and is equivalent to treating all other nuclear spins with a mean field. 

In the generalized CCE (gCCE) formulation, the cluster Hamiltonian includes the full central spin Hamiltonian, allowing Eq.~\ref{eq:clusterHamiltonian} to be rewritten without the projection onto the electron spin states.~\cite{Yang:2020, Onizhuk:2021} Further, instead of expanding the coherence function, one expands the full RDM for the central spin as a product of irreducible contributions,~\cite{Onizhuk:2021b}
\begin{equation}
    \rho_{ij}(t) = \tilde{\rho}_{ij}^{\{0\}} \prod_i \tilde{\rho}_{ij}^{\{i\}} \prod_{i,j} \tilde{\rho}_{ij}^{\{i,j\}} ...,
\end{equation}
enabling gCCE to access population dynamics. 

\begin{figure}[ht!]
    \centering
    \includegraphics[width=\linewidth]{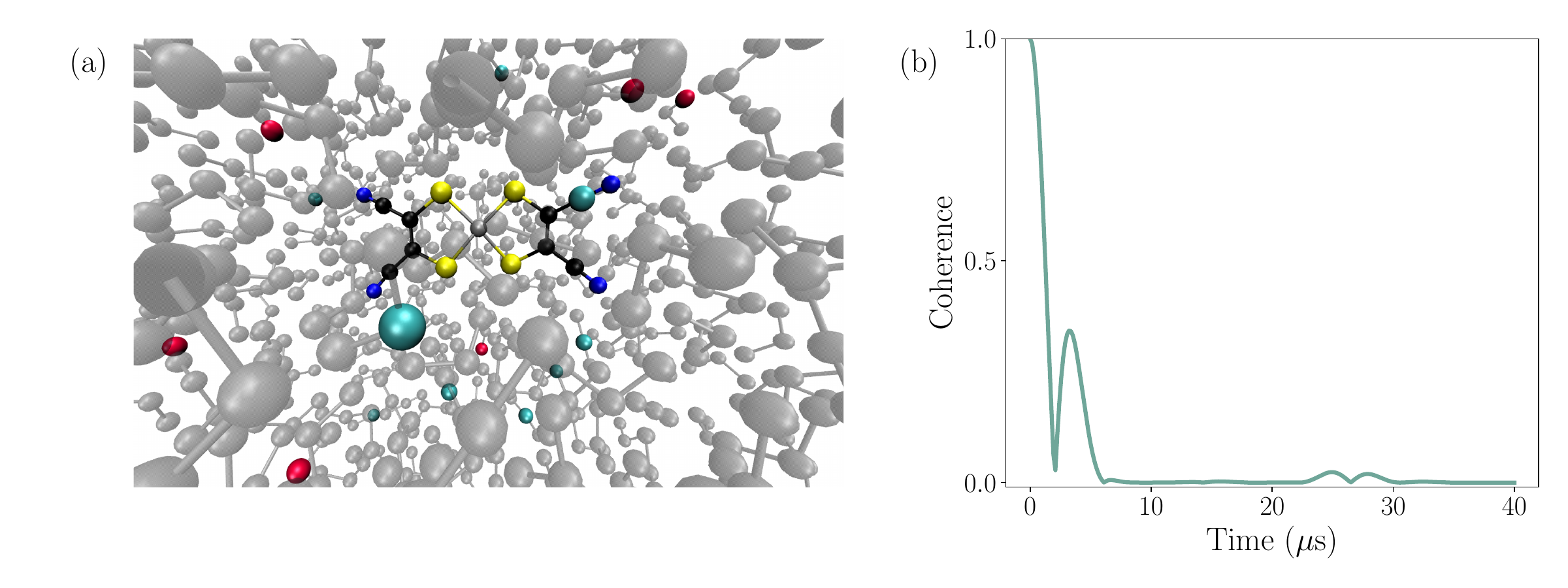}
    \caption{(a) Representation of [Ni(mnt)$_2$]$^-$ molecule solvated in CS$_2$ where cyan indicates $^{13}$C atoms and red indicates $^{33}$S atoms in the bath. Among the spin-free atoms, yellow indicates $^{32}$S, black indicates $^{12}$C and silver indicates the $^{58}$Ni center. (b) Coherence dynamics using gCCE-3.}
    \label{fig:nimnt}
\end{figure}

Both CCE and gCCE have been successfully applied to benchmark decoherence in a wide range of molecular spin systems.~\cite{Chen:2020, Kanai:2022, Ryan:2025} Here, we consider a representative example where CCE succeeds in simulating convergent and numerically stable dynamics: a nickel ion complex, [Ni(mnt)$_2$]$^-$ (mnt2- = maleonitrile-1,2-dithiolate), solvated in CS$_2$.~\cite{Bader:2017} The spin bath here consists of 26 spins: 4~$^{14}$N atoms from the Nickel complex and 13~$^{13}$C and 9~$^{33}$S included in accordance with their natural isotopic abundances (1.1\% and 0.76\%, respectively), as depicted in Figure~\ref{fig:nimnt}~(a). Figure~\ref{fig:nimnt}~(b) illustrates that gCCE-3 yields numerically stable dynamics, where the coherence decays over around $\sim10$\textmu s, which agrees with experimentally observed timescales.~\cite{Bader:2017} gCCE has also been used to model spin dynamics in a variety of other molecular systems, capturing accurate experimental timescales and mechanisms.~\cite{Chen:2020,Jahn:2024,Jahn:2022,Canarie:2020}

Another factorization approach is the analytical pair product approximation (APPA), in which the exact density matrix $\rho(t)$ is solved analytically for a simplified spin Hamiltonian.~\cite{Jeschke:2022} In the limit that the external magnetic field is large, $\hat{H}_{\textrm{spin}}$ can be limited to include only the $A^{zz}$ component of the hyperfine interaction and the dipolar coupling between nuclear spins. The resulting Hamiltonian is often called the secular spin Hamiltonian,
\begin{equation}
    \hat{H}_{\textrm{sec}} = \mu_Bg_eB_z\hat{S}_z+\sum_n\mu_Ng_NB_z\hat{I}_{n}^z + \hat{S}^z\sum_nA_n^{zz}\hat{I}_{n}^z + \sum_{n\neq m}b_{nm}\bigg( \hat{I}_{n}^z\hat{I}_{m}^z - \frac{1}{4}( \hat{I}_{n}^+\hat{I}_{m}^- + \hat{I}_{n}^-\hat{I}_{m}^+ ) \bigg)
    \label{eq:secularSpinHamiltonian}
\end{equation}
where $b_{nm}$ is the dipolar coupling constant between nuclear spins $n$ and $m$. $\hat{H}_{\textrm{sec}}$ induces flip-flops between nuclear spins, which cause pure dephasing of the electron. In the APPA, Eq.~\ref{eq:secularSpinHamiltonian} is taken to include only the doublet electron spin and two nuclear spin-$\frac{1}{2}$ particles. The propagated density matrix in the presence of a single Hahn-echo pulse can be solved exactly by performing the matrix multiplication in Eq.~\ref{eq:unitary-evol} and performing a partial trace over the two nuclear spin degrees of freedom to obtain,
\begin{equation}
    \rho_{q,nm}^{01}(t) = \frac{1}{2}-2\bigg( \frac{b_{nm}\Delta_{nm}}{b_{nm}^2+\Delta_{nm}^2} \bigg)^2\textrm{sin}^4\bigg( \frac{t}{4}\sqrt{\Delta_{nm}^2+b_{nm}^2} \bigg),
    \label{eq:APPA}
\end{equation}
where $\Delta_{nm}=A_n^{zz}-A_m^{zz}$. The many body solution is taken to be the product of all pairwise nuclear spin interactions, 
\begin{equation}
    \rho_q^{01}(t) = \Pi_{n\neq m}\rho_{q,nm}^{01}(t).
    \label{eq:APPAmanyBody}
\end{equation}
The product structure is similar to CCE-2, captures the flip-flop mechanism, and has been successfully applied to a selection of molecular spin qubits.~\cite{Suchaneck:2026}

Despite the success of these factorization approaches, recent work has shown that these approximations tend to overestimate the rate of spin relaxation as a result of the multiplicative nature.~\cite{Li-SBt:2026, Ryan:2026} Additionally, in dense spin baths where many-body correlations become important, factorization approaches struggle to accurately compute dynamics. Further, while factorization approaches tame the dimensionality problem for large spin systems, especially in the Gaussian limit, they face similar challenges as exact diagonalization when including phonon contributions to decoherence, since the bosonic degrees of freedom still require a truncation and systematic convergence of an infinitely large basis. Alternatively, one may combine these with hybrid approaches, such as those highlighted in Sec.~\ref{ssec:mes}. Therefore, while factorization can capture important dephasing dynamics and relaxation effects in specific parameter regimes, alternative methods are necessary.

\subsection{Matrix Product State Approaches}

Matrix product state (MPS) approaches are another class of methods to simulate the dynamics of many-spin systems.~\cite{Banuls:2023, Schollwock:2011, Chan:2012} Motivated by the computational cost of exact diagonalization, low-rank representations of quantum states offer an alternative path to simulating significantly larger spin systems. MPS representations use a sequence of low-rank tensors to represent quantum many-body states and operators.~\cite{Verstraete:2008, Orus:2014, LiMolecular:2026} However, growing entanglement of the system, for example due to the increasing connectedness of the Hamiltonian, generally raises the cost of MPS methods, making the simulation of interacting spin networks with intra-bath interactions challenging. Previous MPS-based approaches to evaluating spin dynamics have focused on central spin-type models where intra-bath interactions are neglected.~\cite{Lindoy:2018}

The Spin-Bath Truncated MPS (SB-tMPS) is a numerically exact and uniformly convergent method to predict the coherence dynamics of interacting many-spin systems. The SB-tMPS method evolves the density matrix of the many-spin system using the MPS formulation while taming the cost of simulating real systems by leveraging physical properties.~\cite{Li-SBt:2026} The density matrix of the system can be represented by an MPS of the form,
\begin{equation}
    \rho_{i,j,k_1,l_1,\cdots,k_N,l_N} = \sum_{\{r_1,\dots,r_{2N+1}\}} W_1(i,r_1)W_2(r_1,j,r_2) \cdots  W_{2N+1}(r_{2N},k_N,r_{2N+1})W_{2N+2}(r_{2N+1},l_N),
\end{equation}
where $i$ and $j$ label the system degrees of freedom, and $k_i$ and $l_i$ index the row and column degrees of freedom of the $i$th bath spin. The tensors $W$ are three-dimensional, encoding the correlations between row and column indices and neighboring sites in the MPS representation. Assuming an initial product state, $\rho_0 \otimes \rho_1 ... \otimes \rho_N$, the density matrix is Choi-transformed into Liouville space to half the length of the MPS chain with the following representation,~\cite{Li:2023}
\begin{equation}
\label{eq:mps}
     |\rho_{i,j,k_1,l_1,\cdots,k_N,l_N}\rangle\rangle = \sum_{\{r_1,...,r_N\}} \tilde W_1(ij,r_1)\tilde W_2(r_1,k_1l_1,r_2) \tilde W_{N+1}(r_N,k_Nl_N),
\end{equation}
where $| \cdot \rangle \rangle$ is the vectorized density matrix. Matrix Product Operators (MPOs) $O^{k}$ are constructed via direct sums of one- and two-site operators to represent the action of $-i\mathcal L \rho$, represented in Liouville space,
\begin{equation}
    O^{(k)}_{ij,\bar i\bar j,k_1l_1,\bar k_1\bar l_1,\cdots,k_Nl_N,\bar k_N\bar l_N} = \sum_{\{r_1,\ldots,r_N\}}
    \tilde M_1(ij,\bar i\bar j,r_1)
    \tilde M_2(r_1,k_1l_1,\bar k_1\bar l_1,r_2) \cdots \tilde M_{N+1}(r_N,k_Nl_N,\bar k_N\bar l_N).
\end{equation} 
Coupling strengths in spin Hamiltonians can have large separations in energy scales. For example, the strength of dipolar interactions decays inversely with the distance cubed, making long-range interactions significantly weaker than short-range interactions. The total Hamiltonian for this system can then be recast in terms of strong and weak-coupling terms,
\begin{align}
    \hat{H}_\mathrm{T} = \sum_{k=1}^{K'} \hat{h}_k \;+\; \sum_{k=K'+1}^{K} \hat{h}_k^{W},
\end{align}
where $h_k =\prod_i\bigotimes_i h_{k,i}$ collects all one-site terms and strongly coupled two-site interactions, and $h_k^{W} =\prod_i\bigotimes_i h^W_{k,i}$ collects weakly coupled terms, including intra-bath or long-range interactions. This physically informs a singular value decomposition (SVD) threshold for various interactions in the MPO. In commonly used quantum sensing systems such as molecular qubits and nitrogen-vacancy centers, the ratio of the interaction strengths $|A_{ij}|/|J_{ij}| \sim 10^3$, allowing for truncation of the singular values of the intra-bath interactions.~\cite{Mohr:2025, Levitt:2008} These innovations make the SB-tMPS method stable and uniformly convergent, allowing this approach to address a wide range of spin qubits.~\cite{Li-SBt:2026}

\begin{figure}[h!]
    \centering
    \includegraphics[width=\linewidth]{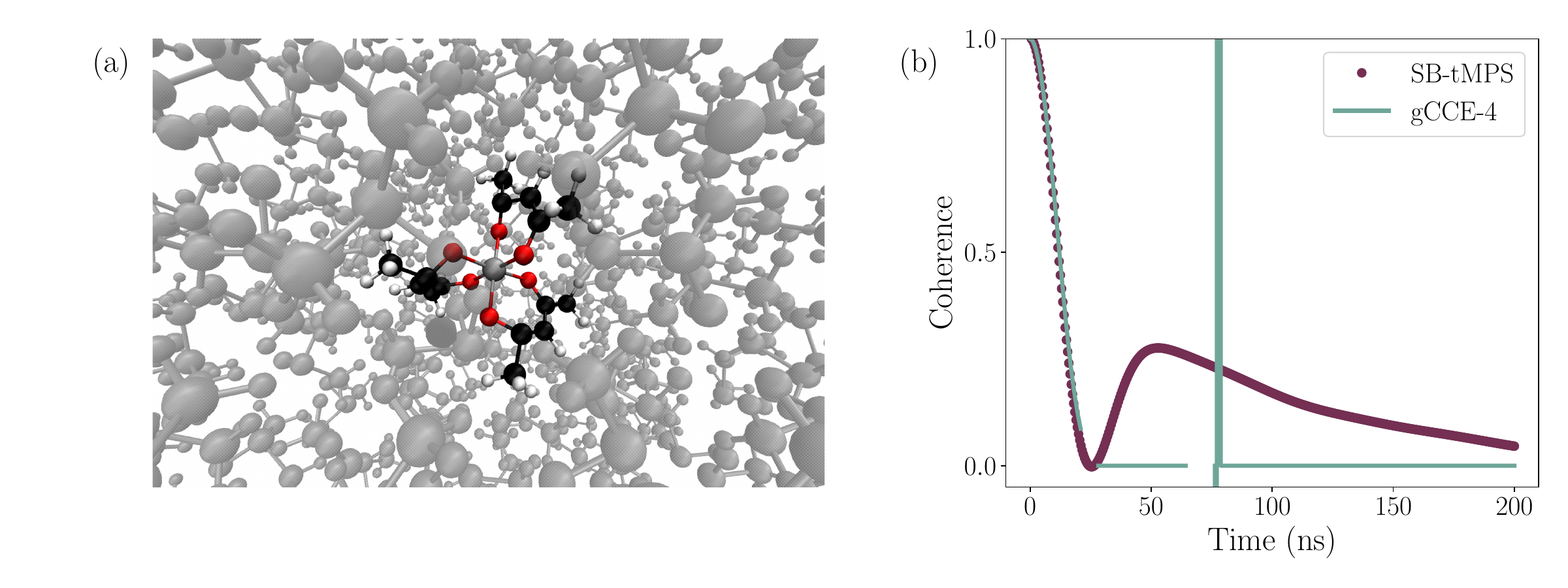}
    \caption{(a) Representation of Ti(acac)$_3$ solvated in DMSO where red indicates oxygen, black indicates carbon and silver indicates titanium. (b) Coherence dynamics using gCCE-4 and SB-tMPS methods. Points with missing CCE data are due to divisions by zero.}
    \label{fig:tiacac}
\end{figure}

As an example, we consider the coherence dynamics of a titanium complex, Ti(acac)$_3$, solvated in DMSO.~\cite{Yu:2016} For this system, the bath consists of 88 spins: 43~$^1$H atoms from the titanium complex and an additional 45~$^1$H atoms from solvent molecules, as shown in Figure~\ref{fig:tiacac}~(a). In Fig.~\ref{fig:tiacac}~(b), we compared the dynamics of this system using the SB-tMPS and CCE methods, demonstrating stable SB-tMPS results and numerical instabilities in the factorization method. 

Tensor network approaches have had enormous success in simulating the quantum dynamics of open quantum systems,~\cite{Shi:2018, Tamascelli:2019, Guan:2024, Keeling:2026} particularly in the paradigmatic spin-boson and related models, and in the context of studying decoherence resulting from an effective bosonic or Gaussian environment.~\cite{Chen:2026, Ye:2026} However, the spin-boson model does not capture Raman-like two-phonon effects or non-Gaussian environments. These two-phonon terms pose a challenge to traditional tensor network-based, as well as other numerically exact, approaches for open quantum systems. While tensor network simulations have demonstrated their ability to simulate both spin baths and phonon baths, simulating quantum dynamics in the presence of both interactions has proven difficult due to the large dimensionality of the problem.

\subsection{Master Equations}
\label{ssec:mes}

Master equations and their generalized counterparts (GMEs) offer a low-dimensional equation of motion for the variables of interest, such as a qubit's reduced density matrix, in a large, potentially infinite open quantum system.~\cite{Nakajima:1958, Zwanzig:1960, Mori:1965, Shibata:1977, Chaturvedi:1979} Master equations that account for non-Markovianity, or memory effects, are denoted as GMEs. GMEs offer an exact rewriting of a dynamics problem that gives explicit access only to select degrees of freedom as well as a starting point for approximations. In QIS, MEs and GMEs have traditionally been combined with low-order perturbation theories to tame the computational complexity of the many-spin problem while offering an efficient route to dynamics predictions. In this (G)ME formalism, nuclear spin environments and thermal lattice vibrations constitute the bath, with longitudinal and transverse decoherence governed by environmental correlation functions.~\footnote{In these (G)MEs, the Argyres-Kelley projector is often adopted,~\cite{Argyres:1964} meaning that the (G)ME offers an equation of motion for the qubit's reduced density matrix, including populations and coherences. This implies that the bath is projected out. For this reason, the literature on these (G)MEs often conflate the bath with the subspace that is projected out; however, this need not be the case, as has been shown in numerous (G)MEs.~\cite{Sparoaglione:2002, Sayer:2026, Mu:2026, Bhattacharyya:2024, Fischer:2007, Reichman:2005, Grabert:2006, Montoya-Castillo:2017, Ferraro:2008, Ferraro:2010, Ayaz:2022, Vroylandt:2022, Mori:1965}} Whether these processes are Markovian or require non-Markovian effects depends on the memory time of these correlation functions relative to the system dynamics. At low temperatures, where local nuclear spin bath fluctuations evolve on a timescale much slower than electron spin dynamics, the system requires a non-Markovian treatment.

In the simpler case of non-interacting spin baths, where spin-spin interactions due to $H_{\rm ss}$ in Eq.~\ref{eq:Hs} are neglected, and in the absence of spin-phonon effects, MEs have successfully modeled spin relaxation in quantum dots and radical pair relaxation mechanisms.~\cite{Lindoy:2018, Lewis:2014, Lewis:2016, Pazera:2024} However, in molecular spin systems, intra-bath nuclear dipolar couplings often cannot be neglected, as nearby nuclear spins generate local field fluctuations that induce electron dephasing. In addition, non-interacting spin-bath models exhibit unphysical relaxation behavior when subject to dynamical decoupling pulses, even at the spin-echo level.~\cite{Maze:2008, Ramon:2017} To capture these intra-bath interactions, the master-equation cluster-correlation expansion (ME-CCE) approach was developed, extending CCE to include explicit dissipation.~\cite{Onizhuk:2024a} Instead of propagating pure unitary cluster dynamics, the cluster RDM, $\hat{\rho}_A(t)$, evolves via the Lindblad master equation, 
\begin{equation}
    \frac{d}{dt}\rho_A(t) = -i\big[ \hat{H}_A,\rho_A(t) \big]+\sum_k\gamma_k\hat{L}_k\rho_A\hat{L}_k^{\dagger}-\frac{\gamma_k}{2}\big\{ \hat{L}_k^{\dagger}\hat{L}_k,\rho_A \big\},
    \label{eq:lindblad}
\end{equation}
where $\hat{H}_A$ is the cluster Hamiltonian, $\hat{L}_k$ are system jump operators, and $\gamma_k$ are the corresponding relaxation rates. The resulting cluster coherence expectation values, $\bra{0}\rho(t)\ket{1}$, are then included in the product expansion Eq.~\ref{eq:cce-prod-exp}. Use of a Lindbladian formalism implies adoption of lowest-order perturbation theory in the Markovian limit, giving access to relaxation rates.~\cite{Weiss:2012, Breuer:2007} These perturbative rates, $\gamma_k$, which can be taken as phenomenological free parameters or calculated \textit{ab initio}, allow ME-CCE to incorporate relaxation channels and inter-spin environment correlations. For instance, one may leverage Fermi’s golden rule, the lowest-order perturbation theory, for weak spin-phonon coupling, which offers relaxation rates for one-phonon processes; accessing two-phonon processes requires summation over intermediate spin states. Second-order frameworks, including the Redfield ME, can capture $T_1$ and $g$- and hyperfine-tensor fluctuations,~\cite{Dmitriev:2025} whereas fourth-order perturbative expansions explicitly yield two-phonon pure-dephasing mechanisms.~\cite{Lunghi:2026} 

To parametrize MEs directly from electronic structure, recent work computed the dephasing rates in Eq.~\ref{eq:lindblad} by sampling a distribution of molecular orientations and geometries for experimental qubit candidates.~\cite{Graham:2017, Krogmeier:2024} As discussed in Sec.~\ref{sec:context}, this incorporation of static disorder in the spin parameters indicates that spin-phonon coupling is treated assuming nuclear motions that are slow on the timescale of spin-spin relaxation. At low temperatures, the flip-flop of two nuclear spins, driven by dipolar coupling, is a major contributor to electron spin dephasing.~\cite{Yao:2006,Zhao:2012} Each individual flip-flop rate can be calculated as,~\cite{Sousa:2003,Vleck:1948}
\begin{equation}
    T_{nm}=2\sqrt{2\pi}A(I)\frac{J_{nm}^2}{K_{nm}}\textrm{exp}\bigg(-\frac{\Delta^2}{8K_{nm}^2}\bigg),
    \label{eq:nucFlipFlop}
\end{equation}
where $J_{nm}$ is the dipolar coupling between nuclear spins $n$ and $m$, $A(I)$ is a normalization factor, $\Delta$ is the difference in hyperfine coupling strength between nuclear spins $n$ and $m$, and $K_{nm}$ is given by,
\begin{equation}
    K_{nm} = \frac{16}{3}I(I+1)\sum_{p\neq n,m}(J_{np}-J_{mp})^2.
\end{equation}
$\Delta$ represents the detuning of the nuclear spin energy levels and suppresses flip-flops when large enough. The dipolar coupling between the two nuclear spins, $J_{nm}$, as well as the dipolar coupling from other nearby nuclear spins, $K_{nm}$, can provide enough energy to overcome the energy gap and allow flip-flops to occur. $\Delta$, $J_{nm}$, and $K_{nm}$ are all anisotropic and therefore depend on molecular orientation. The dephasing rate $\gamma_k$ for a particular nuclear spin pair is taken to be the standard deviation of the flip-flop rates $T_{nm}$, providing a physically-informed estimate of electron spin dephasing from static disorder in the anisotropic spin-spin interactions. By embedding the \textit{ab initio}-derived distribution of these rates into the ME-CCE2 framework, $\gamma_k$ is transformed from an adjustable parameter into a physically motivated value derived from electronic structure theory. This approach accurately captured the spin diffusion barrier and explained the enhanced coherence of copper–selenium qubits over their sulfur-based analogs.~\cite{Graham:2017, Fataftah:2019,Krogmeier:2024} This theoretical treatment thus provides a mechanistic understanding of and confirms the experimental hypothesis that increased spin delocalization onto selenium ligands prolongs dephasing: greater delocalization strengthens the electron–nuclear coupling, thereby increasing the hyperfine detuning between nuclei and inhibiting the flip-flops that drive decoherence.~\cite{Krogmeier:2024} 

When memory effects become important, one must resort to GMEs. When combined with second-order perturbation theory, the time-convolutionless (TCL) form of the GME~\cite{Chaturvedi:1979} yields the TCL2: one of the most useful theoretical frameworks to describe relaxation in open quantum systems.~\cite{Bloch:1957, Redfield:1957, Breuer:2007, Nitzan:2024} In TCL2, as in other second-order perturbation theories that recover the qubit's reduced density matrix from an initially factorized initial condition, what is needed is the bath autocorrelation function of the bath contribution to the qubit-bath interaction. Because the bath, consisting of the spin bath, its intrabath interactions, and additional spin-phonon coupling, still represents a challenging many-body problem, various TCL2 theories for the molecular qubit relaxation problem must incorporate additional approximations. For example, early work on central spin problems neglected intra-bath interactions to control the computational complexity.~\cite{Fischer:2007, Ferraro:2008, Barnes:2011, Barnes:2012}

Another formulation of a TCL2 equation for molecular spins in the pure dephasing limit, including intra-bath spin interactions and a single Hahn-echo pulse for better experimental comparison, has recently been developed.~\cite{Krogmeier:2026} Spin factorization, similar to CCE and the APPA, yielded a GME relating hyperfine couplings to spin dephasing, offering qualitative experimental accuracy. In this TCL2 approach, the dephasing of an electron spin qubit interacting with two homonuclear spin-$\frac{1}{2}$ nuclei can be written as,
\begin{equation}
    \rho_q^{01}(t) = \rho_q^{01}(0)e^{-\chi(t)},
    \label{Eq:rho01}
\end{equation}
with $\chi(t)$ given by,
\begin{equation}
    \chi(t)=\int_0^tdt\int_0^{t_1}dt_1 C(t,t_1),
    \label{eq:corrFn}
\end{equation}
where $C(t,t_1)$ is the correlation function of the nuclear spin bath. This integral can be solved analytically in the case of two homonuclear spin-$\frac{1}{2}$ nuclei to obtain
\begin{equation}
    \chi_{12}(t) = \bigg(\frac{2\Delta_{12}b_{12}}{\Delta_{12}^2+b_{12}^2}\bigg)^2 \textcolor{black}{ \textrm{sin}^4\bigg( \frac{t}{4}\sqrt{\Delta_{12}^2+b_{12}^2} \bigg)},
    \label{Eq:Wtsolved}
\end{equation}
where $b_{12}$ is the nuclear dipolar coupling and $\Delta_{12}$ hyperfine detuning. We note that Eq.~\ref{Eq:rho01} recapitulates the APPA expression in Eq.~\ref{eq:APPA}, demonstrating that APPA is akin to a TCL2 theory in the pairwise nuclear correlations. 

One can simulate the many-body dynamics within this TCL2 by factorizing the correlation functions; the factorization cluster size can be increased to improve accuracy beyond the pair approximation by integrating over the entire cluster space in Eq.~\ref{eq:corrFn}. For pairs of two nuclear spins, the system coherence is given as,
\begin{equation}
    \rho_q^{01}(t)=\rho_q^{01}(0) e^{-\sum_{kl}\chi_{kl}(t)},
    \label{eq:tclFactorized}
\end{equation}
where $k,l$ index nuclear spins $k$ and $l$. Additionally, the TCL2 equation can be solved for heteronuclear spin pairs such as vanadium-hydrogen, demonstrating that such pairs contribute only negligibly to dephasing at low temperatures. For all types of nuclei, hyperfine couplings can be computed using electronic structure on a single geometry, allowing one to interrogate the impact of different types of electronic structure on electron-spin dephasing.~\cite{Krogmeier:2026} Two distinct advantages of the TCL approach include its embedding of the spin-spin dephasing problem with an interacting bath and a pulse sequence into an ME framework, and the ability to include higher orders of perturbation to improve the accuracy of the dynamics. 
\begin{figure}[h!]
    \centering
    \includegraphics[width=0.75\linewidth, trim = 0cm 0cm 13cm 0cm, clip]{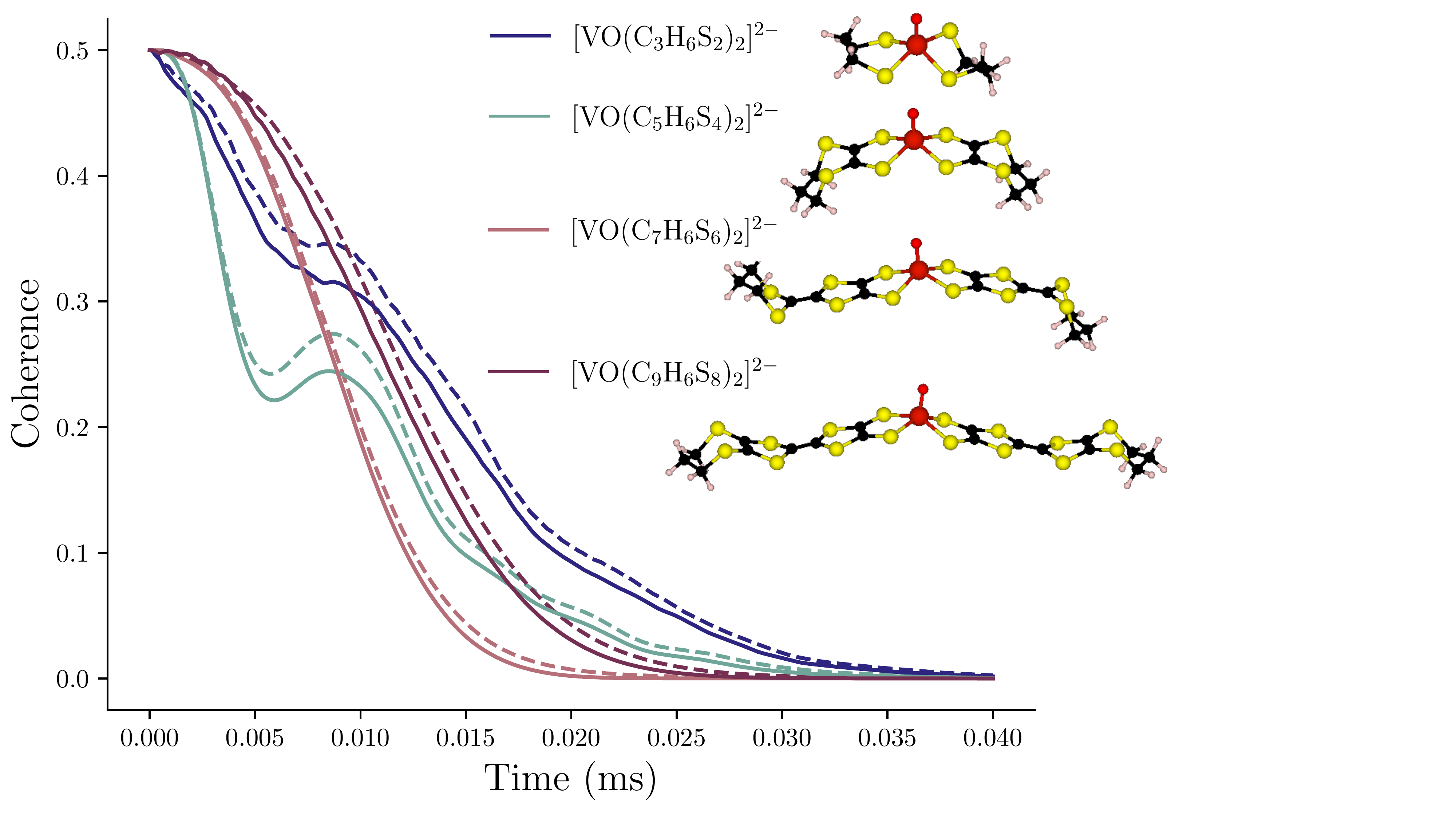}
    \caption{Dephasing dynamics in vanadium molecular series, where the hydrogen spins causing dephasing get further from the electronic state housed on the vanadium center as the molecular size increases. The solid lines are the result of APPA calculations, while the dashed lines are from the TCL2 approach. Vanadium is shown in dark red, oxygen in red, sulphur in yellow, carbon in black, and hydrogen in white.}
    \label{fig:vanadium-example-TCL-APPA}
\end{figure}
Notably, the TCL framework enables the simulation of large molecular environments with hundreds of nuclear spins at low computational overhead. As illustrated in Figure~\ref{fig:vanadium-example-TCL-APPA}, TCL2 and APPA accurately reproduce experimental decoherence timescales for a series of vanadium-oxo complexes embedded in a bath of 581 proton spins.~\cite{Graham:2017} By linking non-Markovian MEs directly to \emph{ab initio} hyperfine tensors, the TCL ME provides an efficient, physically rigorous path to predicting $T_2$ dynamics across various molecular spin architectures. 

Cluster factorization and pairwise expansion methods transform complex many-body environmental decoherence into interpretable, pair-resolved contribution maps. By identifying the specific spin pairs or vibrational modes that drive information loss, these theoretical diagnostic tools directly inform chemical synthesis. The analytical form of the TCL2 equation for electron spin dephasing was recently applied to a set of structurally analogous nickel and zinc molecular qubits which had  different decoherence dynamics despite their geometric similarities.~\cite{Martins:2026,Krogmeier:2026b} Pair-wise analysis of nuclear spins on each molecule and nearby solvent spins identified which specific nuclear-spin interactions were most detrimental to coherence and allowed for the design of modifications to suppress these pathways.~\cite{Krogmeier:2026b} Integrating such pairwise diagnostic methods with spin-phonon spectral density engineering would provide a clear roadmap for designing improved molecular spin architectures, bridging electronic structure theory with synthetic quantum materials design.

Current state-of-the-art ME methods have seen success in modeling spin relaxation and dephasing in distinct parameter regimes. The Redfield equation and semiclassical MEs have been applied to $T_1$ relaxation from spin-phonon contributions,~\cite{Lunghi:2022, Lunghi:2020b,Reta:2021, Briganti:2021, Lunghi:2019, Lunghi:2026, Evans:2019, Goodwin:2017, Yu:2020, Dmitriev:2025} and the Lindblad and TCL MEs have been applied to $T_\phi$ dephasing.~\cite{Krogmeier:2024, Krogmeier:2026, Krogmeier:2026b} Despite the success of these MEs, their applicability is limited by the underlying Hamiltonians from which they are derived, along with their perturbative assumptions; thus, significant work is needed to generalize MEs to capture spin dynamics in parameter regimes where multiple mechanisms are simultaneously active.

\subsection{Special case: Gaussian limit}

When the description of a qubit interacting with its environment is well modeled by Gaussian statistics, the problem simplifies to capturing the dynamics of the well-studied spin-boson model. Methods to capture the quantum dynamics of the spin-boson model and related models are robust and span the gamut from perturbative to semiclassical and numerically exact.   

Two regime distinctions are worth making. In the pure dephasing limit, the second-order cumulant method captures the dynamics of the problem exactly and semianalytically,~\cite{Mukamel:1995} even when the system undergoes multiple rotations under dynamical decoupling in the impulsive limit.~\cite{Szankowski:2017, Krzywda:2019} Beyond the pure dephasing or impulsive limits, more sophisticated and numerically demanding methods are needed. This includes numerically exact techniques like the hierarhical equations of motion (HEOM),~\cite{Tanimura:1989, Tanimura:2006, Bai:2024} path integral techniques,~\cite{Makri:1995a, Makri:2020, Makri:2020b} the multi-configuration time-dependent Hartree method,~\cite{Meyer:1990, Beck:2000, Wang:2003, Meyer:2012, Worth:2008} pseudomode schemes~\cite{Tamascelli:2018, Park:2024}, tensor network-based methods,~\cite{Shi:2018, Strathearn:2018, Tamascelli:2019, Yan:2021, Fux:2021, Bose:2022, Bai:2024, Chen:2025} among others.\cite{Suess:2014, Keil:2001, Chen:2017, Muhlbacher:2008, Somoza:2019} For example, recent efforts have begun exploring hierarchical equations of motion and tensor-network approaches, such as the Time-Evolving Matrix Product Operator (TEMPO) method, for the decoherence of molecular qubits mapped to spin-boson models.~\cite{Bi:2024, Strathearn:2018}

Thus, the methodological landscape for capturing the $T_1$ and $T_2$ timescales and the underlying decoherence mechanisms that arise from spin-spin and spin-phonon effects is rich. It spans the numerically exact---including exact diagonalization, CCE, and tensor network-based methods---to the approximate---such as perturbative Markovian and generalized MEs. In the pure dephasing and Gaussian limits, the second-order cumulant approximation becomes exact and gives access to the dynamics observed under the application of an arbitrary number of pulses via the filter function formalism. When near the Gaussian limit, CCE offers an accurate tool to capture short-time dynamics for spin-spin interactions where the cluster-based correlation can be truncated at few-body orders; in addition, while CCE can in principle produce dynamics arising from the full decoherence limit and the application of dynamical decoupling pulses, it struggles to capture spin-flipping and becomes increasingly expensive when computing dynamics in the presence of multiple pulses. The tensor network-based SB-tMPS addresses many of these challenges, allowing users to predict the dynamics of many-spin systems in the pure dephasing and full decoherence limits and in the presence of many pulses, but is currently limited to $\sim$100 spins. In contrast, perturbative MEs can efficiently tackle many-spin systems with $\geq$ 100 spins and spin-phonon coupling, but rely on the applicability of weak-coupling approximations. Indeed, when spin-phonon interactions are perturbatively weak, researchers have employed Markovian MEs that offer accurate prediction of the temperature dependence of the relevant timescales for different molecular qubits, with Fermi Golden Rule-parametrized Redfield or Lindblad MEs at second or fourth order sufficing to capture direct and Raman relaxation and dephasing pathways. To date, all perturbative ME implementations make the inter-pulse Markovian approximation when accounting for the application of dynamical decoupling pulses.  If one is instead interested in understanding the microscopic mechanism of decoherence in regimes where the weak-coupling or Markovian assumptions start to break down while in the Gaussian limit, where spin-spin and spin-phonon interactions are mapped to a bilinearly coupled spin-boson model, one may use any of a variety of numerically exact approaches. Thus, one must consider the requirements of the physics under investigation before choosing a dynamical approach.

\section{Remaining Challenges and Outlook}
\label{sec:disc}

Molecular spins are playing an increasingly important role in QIS, especially because of the chemical tunability they offer compared to solid-state analogs. This tunability is central when designing tailored molecular platforms to interrogate features of spin physics and develop applications in information processing and sensing. To bring these promises to fruition, we require a fundamental understanding of how these systems respond to environmental dissipation and external fields. Here, predictive theoretical modeling plays a vital role in understanding these underlying mechanisms from first principles. In this Chapter, we have distilled the current state of theoretical modeling of decoherence in molecular spin qubits, with a focus on how the characterization of static and dynamic properties can shed light on operative relaxation mechanisms. We further identified and evaluated the assumptions on which these methods rely to provide a newcomer and expert in the field alike with an accessible resource for navigating this expanding and increasingly important research enterprise. 

Under structural characterization, we focused on both \emph{ab initio} electronic structure and experiment-derived parameterizations of the many-spin Hamiltonian, including spin-phonon interactions, as each can be expected to engender distinct decoherence pathways that vary as a function of temperature, concentration, and applied static and dynamic fields. Outstanding challenges in this structural characterization include capturing both cheaply and scalably: (i) zero-field splitting; (ii) multireference character of the magnetic manifold of electronically excited states; (iii) spin-phonon couplings when the microscoscopic harmonic approximation of nuclear motions starts to fail, which becomes important in glassy matrices and molecular crystals that exhibit significant overtones; (iv) contributions of general molecular motions in the condensed phase.

Under dynamic characterization, we have highlighted the wide variety of methods that can address many-spin Hamiltonians beyond the Gaussian limit as well as the validity of low-order perturbation theory and non-perturbative methods designed for the spin-boson and related models in the Gaussian limit. Here, we have also made a distinction in methods that address the pure dephasing versus full decoherence limits. In the former, which is dominant when spin-flipping is suppressed---such as under dilute conditions, high applied magnetic fields, and low temperatures---dynamics methods can focus on predicting coherence dynamics in the presence of dynamical decoupling pulses. In the latter, where spin flipping occurs on a similar timescale as dephasing, one must predict the dynamics of both coherences and populations to address relaxation in the full qubit tomography. Among the methods reviewed here, we noted that factorization approaches like CCE and APPA currently work best for the pure dephasing limit, while methods such as SB-tMPS and ME-based approaches further extend our ability to simulate spin dynamics in the full decoherence limit. We have also shown that the second-order cumulant---which is exact in the Gaussian, pure dephasing limit---is the only method that can efficiently predict the coherence dynamics under the influence of arbitrary dynamical decoupling protocols. Other methods that can address the non-Gaussian and/or full decoherence limits become increasingly expensive with the size of the system and the number of interacting light pulses. Including spin-phonon couplings remains a challenge to all methods, except perturbative MEs, which are primarily valid in the weak-coupling limit. Thus, outstanding challenges for dynamics methods include accurate prediction of: (i) a large number of both spin and phonon degrees of freedom in the non-Gaussian limit, (ii) arbitrary dynamical decoupling pulse sequences.

The ultimate goal of theoretical modeling in molecular quantum technology is to establish predictive, rational molecular design principles, moving the field from descriptive tools toward predictive discovery of candidates with tailored coherence times and high-fidelity multi-qubit gate operation. The computational frameworks highlighted in this Chapter represent significant strides toward this target. 

\section{Acknowledgements}

K.H.M. acknowledges start-up funding from the University of Minnesota. This work used the computational resources provided by the Minnesota Supercomputing Institute (MSI) at the University of Minnesota. A.M.C.~acknowledges funding from the National Science Foundation (Grant No. 2412615). A.M.C.~also acknowledges the support from a David and Lucile Packard Fellowship for Science and Engineering. This work utilized the Alpine high-performance computing resource at the University of Colorado Boulder. Alpine is jointly funded by the University of Colorado Boulder, the University of Colorado Anschutz, Colorado State University, and the National Science Foundation (Award No. 2201538). This work used the Anvil cluster at Purdue University through allocation CHE250160 from the Advanced Cyberinfrastructure Coordination Ecosystem: Services and Support (ACCESS) program~\cite{Song:2022, Brown:2021}.

\renewcommand*{\bibfont}{\normalsize}
\bibliography{main}

\end{document}